\documentclass[fleqn,usenatbib,onecolumn]{mnras}

\usepackage{newtxtext,newtxmath}

\usepackage[T1]{fontenc}

\usepackage{graphicx}	
\usepackage{amsmath}	

\usepackage{bbold} 

\usepackage{xcolor}

\usepackage{MnSymbol}

\newcommand{\bx}{\mathbfit{x}}
\newcommand{\bk}{\mathbfit{k}}
\newcommand{\bq}{\mathbfit{q}}

\newcommand{\dd}{{\rm d}}
\newcommand{\nn}{\nonumber}

\newcommand{\los}{\hat{n}}

\newcommand{\diff}{\mathrm{d}}
\newcommand{\y}{\mathcal{Y}}

\newcommand{\fnl}{f_\mathrm{NL}}
\newcommand{\Mpc}{\mathrm{Mpc}}

\newcommand{\wtj}[6]{\begin{pmatrix}
   #1 & #2 & #3 \\
   #4 & #5 & #6 
\end{pmatrix}}

\usepackage{ae,aecompl}

\usepackage{newtxtext,newtxmath}

\title[Modal bispectrum]{Modal compression of the redshift-space galaxy bispectrum}

\author[J. Byun and E. Krause]{
Joyce Byun$^{1}$\thanks{E-mail: joyce.byun@protonmail.com}
and Elisabeth Krause$^{1}$
\\
$^{1}$Department of Astronomy and Steward Observatory, University of Arizona, 933 North Cherry Ave, Tucson, Arizona 85721, USA\\
}

\date{Accepted XXX. Received YYY; in original form ZZZ}

\pubyear{2022}

\begin{document}
\label{firstpage}
\pagerange{\pageref{firstpage}--\pageref{lastpage}}
\maketitle

\begin{abstract}
We extend the modal decomposition method, previously applied to compress the information in the real-space bispectrum, to the anisotropic redshift-space galaxy bispectrum.
In the modal method approach, the bispectrum is expanded on a basis of smooth functions of triangles and their orientations, such that a set of modal expansion coefficients can capture the information in the bispectrum.
We assume a reference survey and compute Fisher forecasts for the compressed modal bispectrum and two other basis decompositions of the redshift-space bispectrum in the literature, one based on (single) spherical harmonics and another based on tripolar spherical harmonics. 
In each case, we compare the forecasted constraints from the compressed statistic with forecasted constraints from the full, uncompressed bispectrum which includes all triangles and orientations.
Our main result is that all three compression methods achieve good recovery of the full information content of the bispectrum, but the modal decomposition approach achieves this the most efficiently: only 14 (42) modal expansion coefficients are necessary to obtain constraints that are within 10 (2) per cent of the full bispectrum result.
The next most efficient decomposition is the one based on tripolar spherical harmonics, while the spherical harmonic multipoles are the least efficient.
\end{abstract}




\section{Introduction}

Large-scale galaxy surveys are able to place constraints on cosmological models by comparing the observed positions of galaxies with theoretical predictions, typically in the form of summary statistics describing the clustering properties of galaxies and matter.
In a universe with nearly Gaussian initial conditions, most of the statistical information on galaxy clustering is captured by the two-point correlation function, or its Fourier transform, the power spectrum.
However, due to non-linear gravitational clustering over time, higher-order statistics like the galaxy bispectrum, the Fourier transform of the three-point correlation function (3PCF), encode new complementary information on a wide range of science goals for upcoming galaxy surveys. 
Including the galaxy bispectrum in clustering analyses is expected to yield stronger constraints on galaxy bias and $\Lambda$CDM cosmological model parameters 
\citep{Sefusatti:2006pa,Song:2015gca,Byun:2017fkz,Yankelevich:2018uaz,Gualdi:2020ymf,Agarwal:2020lov,Samushia:2021ixs,Oddo:2021iwq}, 
primordial non-Gaussianity from inflation
\citep{Tellarini:2016sgp,Karagiannis:2018jdt,MoradinezhadDizgah:2020whw,Cabass:2022ymb,Cabass:2022wjy},
neutrino physics
\citep{Ruggeri:2017dda,Hahn:2019zob,Hahn:2020lou,Yankelevich:2022mus},
modified gravity models 
\citep{Yamauchi:2017ibz,Bose:2018zpk,Bose:2019wuz}, 
and relativistic effects at the largest observable scales
\citep{Clarkson:2018dwn,Maartens:2019yhx,deWeerd:2019cae,Maartens:2020jzf}.

Currently, the strongest cosmological constraints from the galaxy bispectrum and 3PCF are from analyses of the SDSS BOSS survey 
\citep{Gil-Marin:2014sta,Gil-Marin:2014baa,Gil-Marin:2016wya,Slepian:2015hca,Gualdi:2018pyw,Philcox:2021kcw}.
While these analyses have included only the monopole component of the anisotropic bispectrum, higher bispectrum multipoles were detected by \citet{Sugiyama:2018yzo} and the (integrated) trispectrum was recently detected by \citet{Gualdi:2022kwz}.
In the future, data from ongoing and future galaxy surveys, such as DESI, Euclid, and SPHEREx, will map the large-scale structure of the Universe in much larger volumes and with higher precision. 
In anticipation of these larger data sets, there is active development on many fronts to improve existing methods, and develop new approaches, for using the galaxy bispectrum as a cosmological observable.

For a standard likelihood-based analysis of galaxy clustering observables, a data covariance matrix is estimated from a large number of realistic mock catalogs generated at a fiducial cosmology.
However, because the galaxy bispectrum is usually measured over a large number of data bins, obtaining accurate covariance matrices this way can be computationally very expensive.
While one way to alleviate the computational burden is to develop techniques to run faster simulations that reduce the computational cost of generating mock catalogs \citep{Colavincenzo:2018cgf}, much effort has also been invested towards finding ways to avoid the brute-force approach to estimating covariance matrices.
In principle, it is possible to theoretically derive an accurate bispectrum covariance matrix and avoid the need for a large number of mocks.
For example, the approach developed by \citet*{Wadekar:2019rdu} to calculate a theoretical covariance matrix for the galaxy power spectrum multipoles may also be promising for the bispectrum.
Another possibility is to estimate the covariance matrix using fewer mocks, for example, by using a shrinkage estimator of the covariance \citep{Joachimi:2016xhk}, fitting a model covariance \citep{Pearson:2015gca}, creating a hybrid covariance matrix using both mocks and theoretical predictions \citep{FriedrichEifler2018,Hall:2018umb}, or using a combination of fast approximate simulations and more expensive full $N$-body simulations \citep{Chartier:2021frd}.

Rather than developing methods for obtaining the covariance matrix, a different strategy is to pursue methods that compress the data vector, such that estimating the corresponding data covariance matrix requires fewer mock catalogs.
This kind of approach can take different forms.
One option is to explore alternative clustering observables that do not try to measure the bispectrum directly, but are still sensitive to higher-order correlations, including the bispectrum.
Some examples of such observables are the line correlation function \citep{Obreschkow:2012yb,Wolstenhulme:2014cla,Eggemeier:2016asq,Byun:2017fkz,Franco:2018yag,Byun:2020hun}, 
the integrated bispectrum \citep{Chiang:2014oga,Chiang:2015eza,Byun:2017fkz},
and skew spectra \citep{Pratten:2011kh,Schmittfull:2014tca,MoradinezhadDizgah:2019xun,Dai:2020adm,Schmittfull:2020hoi}.
The second option is to recover more directly the full cosmological information in the bispectrum through a compressed data set. 
A subset of these methods are general, in that they are not specific to bispectrum analyses, and can also be applied in other contexts, such as the galaxy power spectrum.
These methods generally identify a small number of data bin combinations that contain most of the desired parameter information, and discard the rest of the data. 
For example, \citet{Gualdi:2017iey,Gualdi:2018pyw,Gualdi:2019sfc} compress the bispectrum monopole using the Karhunen-Lo\`eve/MOPED algorithm \citep{Tegmark1997,Heavens:1999am,Alsing:2017var}, and subspace projection also appears to be a promising bispectrum compression method \citep{Philcox:2020zyp}.
Other methods are specifically tailored to bispectrum analyses.
For example, \citet{Gualdi:2019ybt} compress the bispectrum by grouping similar triangles together.
Another approach is to attempt to reconstruct the bispectrum through different basis expansions.
The focus of this work is in this last category.

One of these methods is modal compression, which uses a well-chosen basis to reconstruct the bispectrum.
First developed to search for primordial non-Gaussianity in the CMB \citep{Fergusson:2009nv,Fergusson:2010dm,Ade:2013ydc},
the method has been translated/adapted for LSS clustering \citep{Fergusson:2010ia,Regan:2011zq,Schmittfull:2012hq} and has proven especially fruitful for studying the LSS bispectrum 
\citep{Lazanu:2015rta,Lazanu:2015bqo,Hung:2019ygc,Hung:2019nma}.
Previous work on the real-space matter and halo bispectrum has shown that this method works very well to recover the equivalent constraints that we would get from measuring the standard bispectrum in many triangle bins \citep{Byun:2017fkz,Byun:2020rgl}.

In this work, we extend the modal bispectrum method from real-space to redshift-space, which is a necessary step for developing the modal method to be applied to real galaxy survey data.
Along the way, we have compared it to two other bispectrum compression schemes in the literature, which rely on spherical harmonic \citep{Scoccimarro:2015bla,Rizzo:2022lmh} or tripolar spherical harmonic decompositions of the anisotropic bispectrum \citep{Sugiyama:2018yzo,Sugiyama:2019ike}.
We approach these two other compression schemes as alternative basis expansions and compare their compression efficiency with the modal decomposition.
We calculate Fisher forecasts on the parameters $(b_1, f, \sigma_8, \alpha_\parallel, \alpha_\perp, \fnl)$.
We find that all three compression schemes are able to recover the same forecasted constraints to within 10 per cent of the full anisotropic bispectrum, but the modal decomposition method achieves this the most efficiently, requiring only 32 modal expansion coefficients to achieve constraints that agree to within 2 per cent.

The outline of the rest of this paper is as follows. 
In Section \ref{sec:full_info}, we review the theoretical modeling of the galaxy bispectrum using standard tree-level perturbation theory, and detail how we compute the benchmark Fisher forecast for the full anisotropic redshift-space galaxy bispectrum. 
In Sections \ref{sec:Blm} and \ref{sec:Bl1l2L}, we review the bispectrum multipole decompositions based on spherical harmonics and tripolar spherical harmonics, respectively, and present their corresponding Fisher forecasts.
In Section \ref{sec:modal}, we present the modal decomposition method for the redshift-space bispectrum and present a Fisher forecast.
We conclude with a broader discussion of the main results in Section \ref{sec:discussion}.


\section{The full information content of the redshift-space galaxy bispectrum}
\label{sec:full_info}

In this work, our aim is to compare different ways of recovering the cosmological information contained in the full anisotropic redshift-space bispectrum. 
Our benchmark is the maximum information content that can be recovered by measuring the full anisotropic bispectrum.
We estimate this by computing a Fisher forecast that includes all possible triangle shapes and orientations, without using any compression scheme.
The rest of this section provides the details of how we compute this benchmark Fisher forecast.
We have modeled this forecast loosely on \citet{Gagrani:2016rfy} and our results support the results in this previous work, although the numerical implementation of our calculation is different and we adopt a different reference survey. 

\subsection{Modeling the galaxy power spectrum and bispectrum}
\label{subsec:modeling}

To model the redshift-space galaxy power spectrum and bispectrum, we use the theoretical prediction from standard perturbation theory at tree-level (leading order) \citep{Scoccimarro:1999ed}.
Then the galaxy power spectrum and bispectrum are
\begin{align}
	P_g(\bk) &= Z_1^2(\bk)P_L(k) + P_\mathrm{SN} \label{eq:Ptree} \\
	B_g(\bk_1,\bk_2,\bk_3) &= 2 \, Z_2(\bk_1,\bk_2) Z_1(\bk_1) Z_1(\bk_2) P_L(k_1) P_L(k_2) 
	+ 2\;\mathrm{perms.} + B_\mathrm{SN}(\bk_1,\bk_2,\bk_3)
	\label{eq:Btree}
\end{align}
where $\bk_3 = -(\bk_1+\bk_2)$, $P_L(k)$ is the linear matter power spectrum, and the first- and second-order redshift-space kernels are \citep{Tellarini:2016sgp}
\begin{align}
	Z_1(\bk_1) = \; &(b_1 + f \mu_1^2) + \frac{f_{\mathrm{NL}} b_\phi}{\mathcal{M}(k_1)} \\
	Z_2(\bk_1,\bk_2) = \; &\frac{b_2}{2} + b_1 \left[ F_2(\bk_1,\bk_2) + f_{\mathrm{NL}} \frac{\mathcal{M}(k_3)}{\mathcal{M}(k_1)\mathcal{M}(k_2)} \right]
	+ f \mu_3^2 \left[ G_2(\bk_1,\bk_2) + f_{\mathrm{NL}} \frac{\mathcal{M}(k_3)}{\mathcal{M}(k_1)\mathcal{M}(k_2)} \right] \nonumber \\
	&- \frac{f \mu_3 k_3}{2} \left[ \frac{\mu_1}{k_1} \left(b_1 + f\mu_2^2 + \frac{f_{\mathrm{NL}} b_\phi}{\mathcal{M}(k_2)} \right) 
	+ \frac{\mu_2}{k_2} \left(b_1 + f\mu_1^2 + \frac{f_{\mathrm{NL}} b_\phi}{\mathcal{M}(k_1)} \right) \right] \nonumber \\
	&+ \frac{b_{s^2}}{2} \left( \mu_{12}^2 - \frac{1}{3} \right)
	+ \frac{f_{\mathrm{NL}} b_{\phi\delta}}{2} \left[ \frac{1}{\mathcal{M}(k_1)} + \frac{1}{\mathcal{M}(k_2)} \right] 
	+ f_{\mathrm{NL}} b_{\phi} \frac{\mu_{12}}{2} \left[ \frac{k_2}{k_1}\frac{1}{\mathcal{M}(k_2)} + \frac{k_1}{k_2}\frac{1}{\mathcal{M}(k_1)} \right].
\end{align}
We have defined $\mu_i \equiv \hat{k}_i \cdot \hat{n}$ where $\hat{n}$ is the line-of-sight direction and $\mu_{ij} \equiv \hat{k}_i \cdot \hat{k}_j$.
$F_2$ and $G_2$ are the standard second-order density and velocity kernels \citep{Bernardeau:2001qr}.
$f$ is the growth rate of structure, while $\fnl$ is the amplitude of local-type primordial non-Gaussianity (PNG), and we only include terms that are necessary to describe terms in the power spectrum and bispectrum that are up to first-order in $\fnl$.
$b_1$, $b_2$, and $b_{s^2}$ are the linear, quadratic, and non-local galaxy bias parameters, respectively, while $b_\phi$ and $b_{\phi \delta}$ are two additional PNG galaxy bias parameters.
In this work, we simplify the modeling of galaxy bias by assuming that galaxy bias is local in Lagrangian space, leading to $b_{s^2} = -\frac{4}{7}(b_1-1)$ \citep{Chan:2012jj,Baldauf:2012hs,Saito:2014qha}.
We also assume that the universality relations hold, such that $b_\phi = 2 \delta_c (b_1 - 1)$ and $b_{\phi \delta} = b_\phi - b_1 + 1 + \delta_c [ b_2 - \frac{8}{21}(b_1 - 1)]$ (see \citet{MoradinezhadDizgah:2019xun,Barreira:2021ueb} and the references therein for discussions of the universal mass function assumption).
$\mathcal{M}(k) \equiv 2 k^2 T(k) / 3 \Omega_m H_0^2$, where $T(k)$ is the matter transfer function, relates the linear density fluctuation at late times to the primordial Bardeen potential, $\delta(k) = \mathcal{M}(k) \phi(k)$.

The shot noise contributions are 
\begin{align}
	P_\mathrm{SN} &= \frac{1}{n_g} \\
	B_\mathrm{SN}(\bk_1,\bk_2,\bk_3) &= \frac{1}{n_g} \left[ Z_1^2(\bk_1)P_L(k_1) + 2\;\mathrm{perms.} \right] 
	+ \frac{1}{n_g^2}
\end{align}
where $n_g$ is the average galaxy number density.

In addition to the dynamical redshift-space distortions that are described by the tree-level perturbation theory predictions above, the observed power spectrum and bispectrum are also subject to geometric distortions due to the Alcock-Paczy\'nski (AP) effect \citep{Alcock:1979mp}.
If the assumed cosmology used to translate redshifts and angles into distances and positions differs from the true cosmology, there will be additional distortions parallel and perpendicular to the line of sight.
These geometric distortions can in turn provide clues to the true cosmology.
Here we parametrize the AP effect using two parameters, $\alpha_\parallel$ and $\alpha_\perp$, that distort wave-vectors differently parallel and perpendicular to the line of sight, $\bq_i = \bk_{i\parallel}/\alpha_\parallel + \bk_{i\perp}/\alpha_\perp$, where the $\bk_i$ ($\bq_i$) to correspond to wave-vectors in the assumed (true) cosmology.
Then the observed power spectrum and bispectrum are
\begin{align}
	P_g(\bk) &= \frac{P_g(\bq)}{\alpha_\perp^2 \alpha_\parallel} \\
	B_g(\bk_1,\bk_2,\bk_3) &= \frac{B_g (\bq_1,\bq_2,\bq_3)}{(\alpha_\perp^2 \alpha_\parallel)^2}.
\end{align}
In practice, this means we use the same models for $P_g$ and $B_g$ as in eqs.~\eqref{eq:Ptree} and \eqref{eq:Btree}, but evaluate the wavenumber arguments after AP rescaling, and normalize the power spectrum and bispectrum amplitudes by $\alpha_\perp^2 \alpha_\parallel$ and $(\alpha_\perp^2 \alpha_\parallel)^2$, respectively.

The tree-level perturbation theory modeling that we have summarised here is valid on large scales, while higher-order (loop) contributions and phenomenological Finger-of-God damping factors are often employed to improve the modeling on smaller, more non-linear scales.
In this work, we do not go beyond the tree-level model in eqs.~\eqref{eq:Ptree} and \eqref{eq:Btree}.
While we do not anticipate that more advanced modeling of non-linearities will have a large impact on the relative comparisons between bispectrum estimators, we leave it to future work to explore the impact of non-linearities on the comparison between estimators.

\subsection{Benchmark forecast}

In the benchmark forecast, we calculate a Fisher matrix for the unrealistic scenario where we could measure the galaxy bispectrum for all Fourier-space triangles and orientations without any compression scheme.

If $\delta(\bk)$ is the Fourier transform of the galaxy density contrast field, then it is useful to define the quantity
\begin{equation}
	\hat{\mathcal{B}}(\bk_1,\bk_2,\bk_3) \equiv \frac{\delta(\bk_1) \delta(\bk_2) \delta(\bk_3)}{V} \mathbb{1}_{\bk_{123}}
\end{equation}
where $V$ is the surveyed volume and $\mathbb{1}_{\bk_{123}}$ is equal to one if $\bk_1 + \bk_2 + \bk_3 = 0$ and zero otherwise.
Then $\hat{\mathcal{B}}$ is related to the theoretically predicted bispectrum through ensemble averaging, $\langle \hat{\mathcal{B}} \rangle = B_g$.
The covariance of $\hat{\mathcal{B}}$ in the Gaussian limit is
\begin{align}
	\langle \hat{\mathcal{B}}(\bk_1,\bk_2,\bk_3) \hat{\mathcal{B}}(\bk_1',\bk_2',\bk_3')\rangle 
	= \frac{(2\pi)^9}{V^2} P_g(\bk_1) P_g(\bk_2) P_g(\bk_3)
	&\left[ \delta_D(\bk_1-\bk_1') \Big( \delta_D(\bk_2-\bk_2') \delta_D(\bk_3-\bk_3') 
	+ \delta_D(\bk_2-\bk_3') \delta_D(\bk_3-\bk_2') \Big) \right. \nn \\
	&+ \delta_D(\bk_1-\bk_2') \Big( \delta_D(\bk_2-\bk_3') \delta_D(\bk_3-\bk_1')
 	+ \delta_D(\bk_2-\bk_1') \delta_D(\bk_3-\bk_3') \Big) \nn \\
	&+ \delta_D(\bk_1-\bk_3') \Big( \delta_D(\bk_2-\bk_1') \delta_D(\bk_3-\bk_2')
	+ \delta_D(\bk_2-\bk_2') \delta_D(\bk_3-\bk_1') \left.\!\!\Big) \right] . 
	\label{eq:covBB}
\end{align}
For the vast majority of closed triangles $(\bk_1,\bk_2,\bk_3)$, the covariance will only be non-zero if $(\bk_1',\bk_2',\bk_3')$ describes the same triangle, and only one term out of the six terms on the right side of eq.~\eqref{eq:covBB} will be non-zero.\footnote{
For example, however, an exception to this would be a triangle where $\bk_1 = \bk_2$ and $\bk_3 = - \bk_1 - \bk_2$. In this case, the first and fourth terms on the right side of eq.~\eqref{eq:covBB} would be non-zero.
}
Because of this, we approximate the covariance as a diagonal matrix with
\begin{equation}
	\langle \hat{\mathcal{B}}(\bk_1,\bk_2,\bk_3)\hat{\mathcal{B}}(\bk_1,\bk_2,\bk_3) \rangle = V P_g(\bk_1)P_g(\bk_2)P_g(\bk_3).
\end{equation}

In the continuous limit, the Fisher matrix for the set of parameters corresponding to $\theta_i$ is 
\begin{align}
	\mathbfss{F}_{ij} &= \frac{V}{6} 
	\int \frac{\dd^3 k_1}{(2\pi)^3} 
	\int \frac{\dd^3 k_2}{(2\pi)^3}
	\int \frac{\dd^3 k_3}{(2\pi)^3} 
	(2\pi)^3 \delta_D(\bk_{123})
	\frac{\partial B_g(\bk_1,\bk_2,\bk_3)}{\partial \theta_i} 
	\frac{1}{P_g(\bk_1)P_g(\bk_2)P_g(\bk_3)}
	\frac{\partial B_g(\bk_1,\bk_2,\bk_3)}{\partial \theta_j}.
	\label{eq:full_fisher}
\end{align}
In the last line, the factor of $1/6$ in front of the integrals is necessary because $B_g(\bk_1,\bk_2,\bk_3)$ are equivalent under permutations of the three $\bk_i$ arguments and should not be counted as six separate measurements.

To all forecasts that are presented in this work, we include constraints from the full anisotropic power spectrum. 
The power spectrum Fisher matrix is 
\begin{align}
	\mathbfss{F}^P_{ij} &= \frac{V}{2} \int \frac{\dd^3k}{(2\pi)^3} \frac{\partial P_g(\bk)}{\partial \theta_i} 
	\frac{1}{P_g(\bk)^2}
	\frac{\partial P_g(\bk)}{\partial \theta_j}.
	\label{eq:ps_fisher}
\end{align}

Unless explicitly mentioned otherwise, the forecasts presented in this work do not directly use the continuous integration expressions shown above to compute the Fisher matrices.
Instead, we compute discretized sums over the wave-vectors $\bk_i$ as determined by Fourier-space FFT grids by making the replacement
\begin{equation}
	\int \frac{\dd^3k}{(2\pi)^3} \rightarrow \frac{1}{V} \sum_{\bk}
\end{equation}
in eqs.~\eqref{eq:full_fisher} and \eqref{eq:ps_fisher}.\footnote{For FFT calculations in this work, we use the \href{https://www.intel.com/content/www/us/en/developer/tools/oneapi/onemkl.html}{Intel oneAPI Math Kernel Library}.
}
For the bispectrum Fisher matrix in eq.~\eqref{eq:full_fisher}, the discretized sum is computed efficiently if the partial derivative $\partial B_g/\partial \theta_i$ can be written as a function that is separable in its dependence on $\bk_1$, $\bk_2$, and $\bk_3$.
This is made possible for the bispectrum model and Fisher parameters used in this work by using the modal basis functions that are presented later in Section \ref{subsec:custom_modes}.

\subsection{Fisher forecast settings}

In this work, our goal is to compare different ways of compressing information in the redshift-space galaxy bispectrum. 
For this purpose, we will compute Fisher forecasted constraints on $\boldsymbol{\theta} = (b_1,b_2,f,\sigma_8,\alpha_\perp,\alpha_{||},f_\mathrm{NL})$ for the full redshift-space bispectrum, the spherical harmonic $B_{\ell m}(k_1,k_2,k_3)$ multipoles \citep{Scoccimarro:2015bla}, the tripolar spherical harmonic (TriPoSH) $B_{\ell_1 \ell_2 L}(k_1,k_2)$ multipoles \citep{Sugiyama:2018yzo}, and the modal bispectrum coefficients $\beta_n$.
Constraints from the full redshift-space bispectrum are the benchmark against which the other estimators are measured.

We will assume a reference survey of a single redshift slice with a volume, galaxy number density, and fiducial parameter values that are roughly similar to what will be included as one redshift bin in ongoing and upcoming spectroscopic galaxy surveys such as Euclid, DESI, and SPHEREx. 
More realistic forecasting for a specific survey scenario is outside the scope of this work.
We expect that the forecasts here will still give us a reasonably accurate comparison between compression methods, although the absolute strength of the constraints will differ, but this should be checked for more realistic forecasts across different number densities.

We set our survey volume to be $V = 4.0 \, h^{-3} \, \mathrm{Gpc}^3$, which is approximately representative of a redshift shell at $z=1$ with thickness $\Delta z=0.1$ and a sky fraction of $f_\mathrm{sky} = 0.35$.
The galaxy number density is set to $n_g = 6 \times 10^{-4} \, h^3 \, \mathrm{Mpc}^{-3}$. 
This is similar to what is expected for the Euclid H$\alpha$ and DESI ELG samples at $z \sim 1$ \citep{Euclid:2019clj,DESI:2016fyo}. 
SPHEREx plans to use five different tracers in each redshift bin \citep{Dore:2014cca}, and this number density is roughly what is expected for the tracer with the highest number density.

We use the following cosmological parameter values to generate the linear matter power spectrum and matter transfer function at redshift $z=1$ using CLASS\footnote{\url{http://class-code.net}} \citep{Blas:2011rf}: $\Omega_\mathrm{c} = 0.2642$, $\Omega_\mathrm{b} = 0.0493$, $H_0 = 67.4 \, \mathrm{km} \, \mathrm{s}^{-1} \, \mathrm{Mpc}^{-1}$, $n_s = 0.965$, and $\sigma_8 = 0.811$.
These parameters are consistent with constraints on the flat $\Lambda$CDM cosmological model from the Planck 2018 analysis of CMB temperature and polarization anisotropies and CMB lensing \citep{Planck:2018vyg}.

The fiducial values of the AP parameters are $\alpha_\perp = \alpha_{||} = 1$, which corresponds to the assumption that the fiducial cosmology is the same as the true one.
The fiducial value of $f_\mathrm{NL}$ is zero, as non-zero local PNG has yet to be detected.
The fiducial bias parameters are $b_1 = 1.50$ and for $b_2$ we use the fitting formula for quadratic bias in \cite{Lazeyras:2015lgp}, $b_2 = 0.412 - 2.143 \, b_1 + 0.929 \, b_1^2 + 0.008 \, b_1^3$.

\section{Spherical harmonic decomposition}
\label{sec:Blm}

In this section, we review the spherical harmonic multipole decomposition from \citet{Scoccimarro:2015bla}, compute Fisher forecasts for the $B_{\ell m}$ multipoles, and compare the constraints from $B_{\ell m}$ to the benchmark forecast.
The work in this section obtains results that are similar to \citet{Gagrani:2016rfy}.

\subsection{Definition of $B_{\ell m}$ multipoles}

The redshift-space bispectrum is a function that depends on the shape of the triangle that is formed by $(k_1,k_2,k_3)$, as well as the relative orientation between this triangle and the line of sight, which can be parametrized with two angles.
The $B_{\ell m}$ multipoles decompose this dependence on triangle orientation into spherical harmonics,\begin{align}
	B_{\ell m}(k_1,k_2,k_3) &= \int \frac{\dd^2\hat{n}}{4\pi} B(\bk_1,\bk_2,\los) Y_{\ell m}^*(\theta,\phi),
	\label{eq:Blm_def}
\end{align}
where the angles $(\theta,\phi)$ correspond to the polar and azimuthal angles describing the line of sight direction $\los$ in a coordinate system determined by $\hat{k}_1$ and $\hat{k}_2$, such that $\hat{k}_1$ defines the $z$-axis and $\hat{k}_2$ lies in the $xz$-plane.
We choose the $Y_{\ell m}$ to be normalized such that $Y_{00} = 1$ and
\begin{equation}
	\int \frac{\dd^2\hat{n}}{4\pi} \, Y_{\ell m}(\los) Y^*_{\ell' m'}(\los) = \delta^K_{\ell \ell'} \delta^K_{m m'}. \label{eq:Ylm}
\end{equation}
Since the $B_{\ell m}$ are the multipoles for a real-valued quantity, the multipoles with $m=0$ will be real-valued, and the multipoles with $m \neq 0$ will be complex-valued.
The multipoles for $m<0$ are then $B_\ell^{-|m|} = (-1)^{|m|} B_\ell^{|m|*}$, so we only keep the unique multipoles with $m \geq 0$. 

\subsection{$\hat{B}_{\ell m}$ estimator and covariance}

In the global plane-parallel limit, where the line of sight over the survey volume is fixed to $\hat{\mathbfit{z}}$, the estimator is
\begin{align}
	\hat{B}_{\ell m}(k_1,k_2,k_3) &= \frac{1}{N_\mathrm{tri}} 
	\int_{k_1} \frac{\dd^3q_1}{(2\pi)^3}
	\int_{k_2} \frac{\dd^3q_2}{(2\pi)^3} 
	\int_{k_3} \frac{\dd^3q_3}{(2\pi)^3} (2\pi)^3 \delta_D(\bq_{123})
	\hat{\mathcal{B}}(\bq_1,\bq_2,\bq_3)
	Y_{\ell m}^*(\theta,\phi), 
	\label{eq:Blm_estimator}
\end{align}
where the normalization factor is
\begin{equation}
	N_\mathrm{tri} = \int_{k_1} \frac{\dd^3q_1}{(2\pi)^3} \int_{k_2} \frac{\dd^3q_2}{(2\pi)^3} \int_{k_3} \frac{\dd^3q_3}{(2\pi)^3} (2\pi)^3 \delta_D(\bq_{123}).
\end{equation}
The subscript on the integral, $\int_{k_i}$, is shorthand for indicating that the integral is only over the Fourier-space shell centered at $k_i$.

Then the covariance is
\begin{align}
	&\langle \hat{B}_{\ell m}(k_1,k_2,k_3) \hat{B}^*_{\ell'm'}(k_1',k_2',k_3') \rangle \nn \\
	&= \frac{1}{N_\mathrm{tri}N_\mathrm{tri}'} 
	\int_{k_1} \frac{\dd^3q_1}{(2\pi)^3}
	\int_{k_2} \frac{\dd^3q_2}{(2\pi)^3}
	\int_{k_3} \frac{\dd^3q_3}{(2\pi)^3} (2\pi)^3 \delta_D(\bq_{123})
	\int_{k_1'} \frac{\dd^3q_1'}{(2\pi)^3}
	\int_{k_2'} \frac{\dd^3q_2'}{(2\pi)^3}
	\int_{k_3'} \frac{\dd^3q_3'}{(2\pi)^3} (2\pi)^3 \delta_D(\bq_{123}') \nonumber \\ 
	& \hspace{0.5cm} \times 
	\langle \hat{\mathcal{B}}(\bq_1,\bq_2,\bq_3) \hat{\mathcal{B}}(\bq_1',\bq_2',\bq_3')\rangle
	Y_{\ell m}^*(\theta,\phi) Y_{\ell' m'}(\theta',\phi') \nn \\
	&= \delta^K_{k_1 k_1'} \delta^K_{k_2 k_2'} \delta^K_{k_3 k_3'} 
	\frac{s_\triangle}{N_{tri}^2 V}
	\int_{k_1} \frac{\dd^3q_1}{(2\pi)^3}
	\int_{k_2} \frac{\dd^3q_2}{(2\pi)^3} 
	\int_{k_3} \frac{\dd^3q_3}{(2\pi)^3} (2\pi)^3 \delta_D(\bq_{123})
	P_g(\bq_1)P_g(\bq_2)P_g(\bq_3) 
	Y_{\ell m}^*(\theta,\phi) Y_{\ell' m'}(\theta,\phi)
	\label{eq:Blm_cov_fft}
\end{align}
where we have used eq.~\eqref{eq:covBB} for $\langle \hat{\mathcal{B}} \hat{\mathcal{B}}'\rangle$.
The three Kronecker delta factors emphasize that the covariance is only non-zero if the two triangle bins are the same, $k_1 = k_1'$, $k_2 = k_2'$, and $k_3 = k_3'$.
We note however that generally there is a non-zero covariance between multipoles with different $(\ell,m)$.
$s_\triangle$ is a factor that depends on whether the triangle bin is equilateral ($s_\triangle = 6$), isosceles ($s_\triangle = 2$), or scalene ($s_\triangle = 1$). 

We can simplify this six-dimensional integral expression for the covariance by taking the thin-shell approximation.
Assuming that the width of each $k_i$ bin is small enough that we can safely replace the magnitudes of the $\bq_i$ wave-vectors in the integrand with $k_i$, we can exchange the integral over all triangle orientations for a two-dimensional integral over all $\los$ directions,
\begin{align}
	\langle \hat{B}_{\ell m}(k_1,k_2,k_3) \hat{B}^*_{\ell'm'}(k_1,k_2,k_3) \rangle &= 
	\frac{s_\triangle }{N_\mathrm{tri} V} 
	\int \frac{\dd^2\hat{n}}{4\pi} 
	P_g(\bk_1)P_g(\bk_2)P_g(\bk_3)
	Y_{\ell m}^*(\theta,\phi) Y_{\ell' m}(\theta,\phi).
	\label{eq:Blm_cov}
\end{align}

The bispectrum multipole estimator in eq.~\eqref{eq:Blm_estimator} is usually a computationally infeasible six-dimensional integral over the FFT grids for $\bq_1$ and $\bq_2$. 
\cite{Scoccimarro:2015bla} has noted that this calculation for the $m=0$ multipoles can be computed easily, since $Y_{\ell 0}(\theta,\phi)$ reduces to a Legendre polynomial that only depends on the angle between $\bq_1$ and the line of sight, $Y_{\ell 0}(\theta,\phi) = \mathcal{L}_\ell(\cos\theta_1)$, and the Dirac delta function can also be written in a separable way as
\begin{equation}
	\delta_D(\bq_{123}) = \int \frac{\dd^3x}{(2\pi)^3} e^{i(\bq_1+\bq_2+\bq_3) \cdot \bx}.
	\label{eq:dirac_delta}
\end{equation}
Then the estimator for $\hat{B}_{\ell 0}$ can computed efficiently as
\begin{align}
	\hat{B}_{\ell 0}(k_1,k_2,k_3) &= \frac{1}{N_\mathrm{tri} V} \int \dd^3x
	\left[ \int_{k_1} \frac{\dd^3q_1}{(2\pi)^3} e^{i\bq_1 \cdot \bx} \delta(\bq_1) \mathcal{L}_\ell(\cos\theta_1) \right]
	\left[ \int_{k_2} \frac{\dd^3q_2}{(2\pi)^3} e^{i\bq_2 \cdot \bx} \delta(\bq_2) \right]
	\left[ \int_{k_3} \frac{\dd^3q_3}{(2\pi)^3} e^{i\bq_3 \cdot \bx} \delta(\bq_3) \right].
	\label{eq:Bl0_fft}
\end{align}
However, for the $m \neq 0$ multipoles, it is not clear whether the estimator can be cast into a similarly separable form.
In the forecasted results that follow, we therefore consider two scenarios: one where only the $B_{\ell 0}$ multipoles are included and another where all $B_{\ell m}$ multipoles are included.

\begin{figure*}
\includegraphics[width=\textwidth]{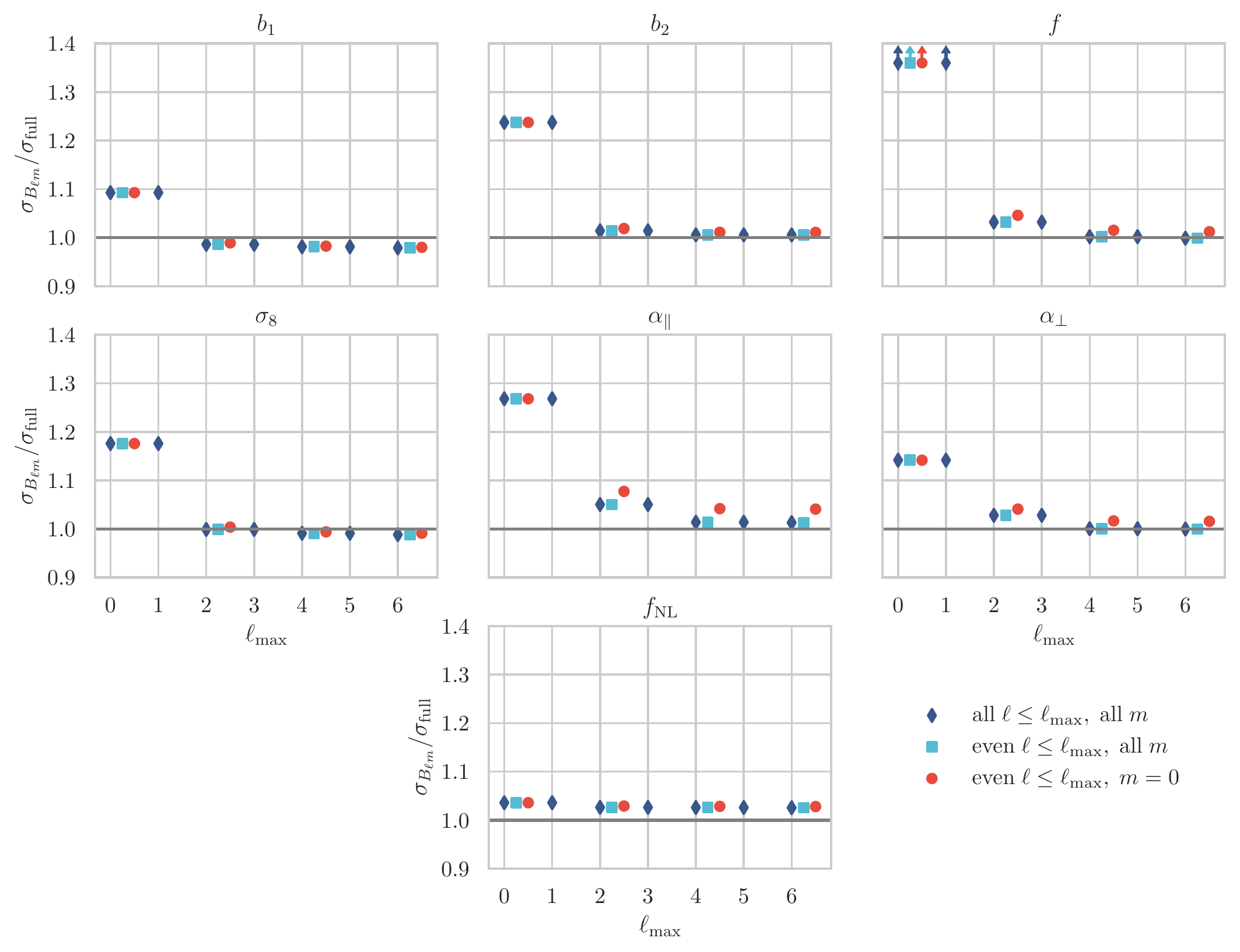}
\caption{Ratio of constraints from $B_{\ell m}$ and the full redshift-space bispectrum, $\sigma_{B_{\ell m}}/\sigma_\mathrm{full}$, as a function of including $B_{\ell m}$ multipoles up to $\ell_\mathrm{max}$ for parameters $(b_1, b_2, f, \sigma_8, \alpha_\parallel, \alpha_\perp, \fnl)$.
We note that for the parameters $b_1$ and $\sigma_8$ in the first column of panels, the forecasted error from the $B_{\ell m}$ multipoles can be up to 3per cent \textit{smaller} than what is forecast from the full bispectrum, due to the thin-shell approximation.
We find that the odd $\ell$ multipoles (included in the dark blue diamonds) do not contribute any additional information, and constraints show negligible improvement by the inclusion of $\ell = 6$ multipoles.
Only a small amount of information is lost by leaving out the $m\neq0$ multipoles, as shown by the difference between light blue squares and red circles.
Constraints up to $\ell_\mathrm{max} = 4$ including all $m \geq 0$ multipoles are within 3 per cent of the full bispectrum constraint. 
This increases very slightly to 5 per cent when the non-zero $m$ multipoles are dropped.
This latter case including the multipoles $(B_{00}, B_{20}, B_{40})$ corresponds to a data vector with 1,197 bins.
}
\label{fig:Blm_fisher}
\end{figure*}

\subsection{$B_{\ell m}$ Fisher forecast results}

The Fisher matrix for $B_{\ell m}$ is 
\begin{equation}
	\mathbfss{F}_{ij} \equiv \sum_{\ell m} \sum_{\ell' m'} \sum_{k_1 k_2 k_3} \frac{\partial B_{\ell m}(k_1,k_2,k_3)}{\partial \theta_i}
	\mathbfss{C}^{-1}[B_{\ell m}(k_1,k_2,k_3), B^*_{\ell' m'}(k_1,k_2,k_3)]
	\frac{\partial B^*_{\ell' m'}(k_1,k_2,k_3)}{\partial \theta_j}
	+ \mathbfss{F}^P_{ij}.
\end{equation}
The $(k_1,k_2,k_3)$ triangle bins are determined by the center of the lowest $k$-bin, $k_\mathrm{min} = 0.02 \, h \, \Mpc^{-1}$, the center of the highest $k$-bin, $k_\mathrm{max} = 0.15 \, h \, \Mpc^{-1}$, and the bin width $\Delta k = 0.01 \, h \, \Mpc^{-1}$.
We only include unique triangle bins by requiring that $k_1 \geq k_2 \geq k_3$.
In total there are 399 triangle bins per multipole.

\begin{figure*}
\includegraphics[width=\textwidth]{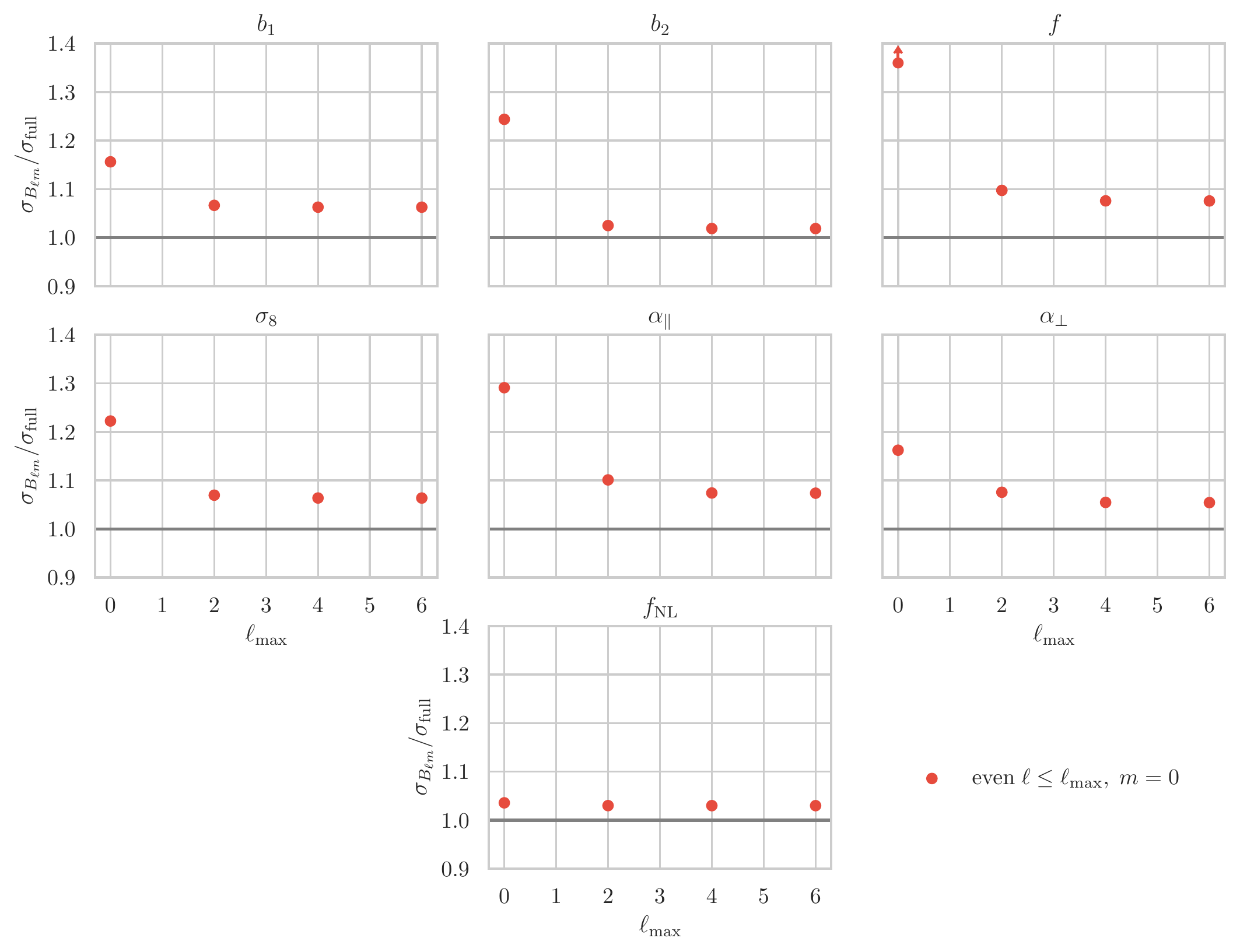}
\caption{The same as Fig.~\ref{fig:Blm_fisher}, except we consider here only the $B_{\ell m}$ multipoles with even $\ell$ and $m=0$ and compute the Fisher matrices using FFT-based expressions that do not assume the thin-shell approximation for $B_{\ell m}$ and its covariance.
We find that constraints converge with $\ell_\mathrm{max}=4$ as in Fig.~\ref{fig:Blm_fisher}, but the agreement between the $B_{\ell0}$ forecast and the full bispectrum constraint loosens slightly from the the 5 per cent in Fig.~\ref{fig:Blm_fisher} to 8 per cent here.
}
\label{fig:Bl0_fisher_fft}
\end{figure*}

Using the continuous integral expressions in eq.~\eqref{eq:Blm_def} to compute the derivatives and eq.~\eqref{eq:Blm_cov} to compute the covariance,\footnote{We used the SHTOOLS library \citep{Wieczorek:2018} available at \url{https://github.com/SHTOOLS/SHTOOLS}.}
 we obtain the Fisher forecasted errors $\sigma_{B_{\ell m}}$ on the parameters $( b_1, b_2, f, \sigma_8, \alpha_\parallel, \alpha_\perp, f_\mathrm{NL} )$.
The errors $\sigma_\mathrm{full}$ from the full redshift-space bispectrum are computed by evaluating eq.~\eqref{eq:full_fisher} as continuous integrals using the CUBA library \citep{Hahn:2004fe,Hahn:2014fua}.\footnote{\url{http://www.feynarts.de/cuba/}}
The ratios between $\sigma_{B_{\ell m}}$ and $\sigma_\mathrm{full}$ are shown in Fig.~\ref{fig:Blm_fisher} for different combinations of multipoles, indicated by three different marker symbols.
We note that in the first column of Fig.~\ref{fig:Blm_fisher}, we find that for $b_1$ and $\sigma_8$, $\sigma_{B_{\ell m}}$ can be up to 3 per cent \textit{smaller} than $\sigma_\mathrm{full}$, which is an unphysical numerical artifact due to the use of the thin-shell approximation in the estimator and covariance expressions in eqs.~\eqref{eq:Blm_def} and \eqref{eq:Blm_cov}.
We have checked that wider bin widths $\Delta k$ exacerbate this discrepancy, while smaller bin widths make the difference smaller, at the cost of rapidly increasing the total number of triangle bins in the forecast.

We identify broad trends in the forecasted constraints that agree with the results in \citet{Gagrani:2016rfy}.
We find that the odd $\ell$ multipoles (included in the points marked by dark blue diamonds) add negligible constraining power.
Including even $\ell$ multipoles up to $\ell_\mathrm{max} = 4$ (as shown by the light blue squares) recovers most of the available information, with constraints on all parameters within 3 per cent of the constraints from the full bispectrum forecast.
If only the $m=0$ multipoles are used (red circles), the constraints are also saturated at $\ell_\mathrm{max} = 4$, but a small amount of information is lost. This is most noticeable for $\alpha_\parallel$, where the error is 5 per cent larger than from the full bispectrum.
This combination of multipoles, $B_{00}$, $B_{20}$, and $B_{40}$, corresponds to a data vector with $399 \times 3 = $ 1,197 bins.

For this last scenario marked by the red circles, including only $B_{\ell 0}$ for even $\ell$, we are able to calculate the Fisher derivatives and covariance matrix using the FFT-based expressions in eqs.~\eqref{eq:Blm_cov_fft} and \eqref{eq:Bl0_fft}.
Similarly, we compute the full bispectrum forecast in eq.~\eqref{eq:full_fisher} using FFTs.
The comparison of FFT-based forecasts is shown in Fig.~\ref{fig:Bl0_fisher_fft}, and we consider this comparison to be more accurate, since it does not use the thin-shell approximation when computing $B_{\ell m}$ and its covariance.
We find that the $B_{\ell 0}$ for $\ell = 0,2$, and $4$ can recover parameter constraints to within 8 per cent of the full bispectrum.


\section{Tripolar spherical harmonic decomposition}
\label{sec:Bl1l2L}

In this section, we focus on a second, different decomposition of the redshift-space bispectrum that uses a basis of tripolar spherical harmonic (TripoSH) functions  \citep{Varshalovich1988} rather than (single) spherical harmonics.
We briefly review the TripoSH decomposition of the bispectrum presented in \citet{Sugiyama:2018yzo,Sugiyama:2019ike}, including its estimator and covariance, before presenting the Fisher forecast for the TripoSH $B_{\ell_1 \ell_2 L}$ multipoles compared to the full redshift-space bispectrum.
To our knowledge, this is the first such comparison in the literature.

We note that the TripoSH basis for the bispectrum is related to many previous works that used bipolar or tripolar spherical harmonics to study the CMB and large-scale structure.
For example, bipolar spherical harmonics were first proposed by \citet{Hajian:2003qq,Hajian:2005jh} to study the statistical isotropy of the CMB and has been used by the Planck collaboration to put constraints on deviations from statistical isotropy \citep{Planck:2015igc}.
For large-scale structure studies, the bipolar spherical harmonics formalism has been used to constrain statistical anisotropy in the galaxy power spectrum \citep{Shiraishi:2016wec,Sugiyama:2017ggb} and tripolar spherical harmonics have also been used to capture wide-angle effects in the two-point correlation function \citep{Szapudi:2004gh}.

\subsection{Definition of TripoSH multipoles}

\citet{Sugiyama:2018yzo} presented an alternative way to decompose the anisotropic bispectrum using a tripolar spherical harmonic (TripoSH) basis. 
The TripoSH basis is a tensor product of three spherical harmonics, $\{\{Y_{\ell_1}(\hat{k} _1) \otimes Y_{\ell_2}(\hat{k}_2) \}_{\ell_{12}} \otimes Y_L(\los)\}_{J M_J}$, that can be used to describe how the bispectrum depends on three directions, $\hat{k}_1$, $\hat{k}_2$ and $\los$.
The advantage of using this basis is that the TripoSH basis functions have many of the same properties as spherical harmonics: they obey similar orthogonality conditions, rotate like spherical harmonics, and behave similarly under coordinate inversions \citep{Varshalovich1988}.
These properties can be used to relate cosmological assumptions to restrictions on the allowed multipole coefficients, which in general would be the full set of $B_{\ell_1 \ell_2 \ell_{12} L}^{J M_J}(k_1,k_2)$ multipole coefficients.
Assuming that the bispectrum is statistically isotropic translates to requiring that $J = M_J = 0$ for the total angular momenta, and it follows that then the only non-zero multipoles are those with $\ell_{12} = L$.
Assuming that the universe is parity symmetric is equivalent to having only the $\ell_1 + \ell_2 + L =$ even multipoles be non-zero.

Under typical assumptions of a homogeneous, isotropic, and parity-symmetric universe then, the bispectrum can be expanded on the TripoSH basis as 
\begin{align}
	B(\bk_1,\bk_2,\los) &= 
	\sum_{\ell_1+\ell_2+L\,=\,\mathrm{even}}
	B_{\ell_1\ell_2L}(k_1,k_2) \;
	S_{\ell_1\ell_2L}(\hat{k}_1,\hat{k}_2,\los),
\end{align}
where the TripoSH basis functions are
\begin{align}
	S_{\ell_1\ell_2L}(\hat{k}_1,\hat{k}_2,\los) &= \frac{1}{H_{\ell_1\ell_2 L}} \sum_{m_1m_2M}
	\wtj{\ell_1}{\ell_2}{L}{m_1}{m_2}{M}
	\y_{\ell_1}^{m_1}(\hat{k}_1) \;
	\y_{\ell_2}^{m_2}(\hat{k}_2) \;
	\y_{L}^{M}(\los)
\end{align}
and the TriPoSH multipoles are
\begin{align}
	B_{\ell_1\ell_2L}(k_1,k_2) &=
	H_{\ell_1 \ell_2 L}
	\sum_{m_1m_2M}
	\wtj{\ell_1}{\ell_2}{L}{m_1}{m_2}{M}
	B_{\ell_1\ell_2L}^{m_1m_2M}(k_1,k_2)
	\label{eq:Bl1l2L} \\
	B_{\ell_1\ell_2L}^{m_1m_2M}(k_1,k_2) &\equiv 
	N_{\ell_1\ell_2L}
	\int \frac{\diff^2 k_1}{4\pi}
	\int \frac{\diff^2 k_2}{4\pi}
	\int \frac{\diff^2 \hat{n}}{4\pi}
	\y_{\ell_1}^{m_1*}(\hat{k}_1) \;
	\y_{\ell_2}^{m_2*}(\hat{k}_2) \;
	\y_{L}^{M*}(\los) \;
	B(\bk_1,\bk_2,\los). \label{eq:intYYYB}
\end{align}
The Wigner 3-$j$ symbols require that $(\ell_1,\ell_2,L)$ combinations satisfy the triangle inequality, $|\ell_1-\ell_2| \leq L \leq \ell_1+\ell_2$.
We use the same notations as in \citet{Sugiyama:2018yzo} and define the factors
$H_{\ell_1 \ell_2 L} \equiv \wtj{\ell_1}{\ell_2}{L}{0}{0}{0}$, which acts to only select multipoles that have $\ell_1 + \ell_2 + L =$ even, and $N_{\ell_1\ell_2L} \equiv (2\ell_1+1)(2\ell_2+1)(2L+1)$.
To match that work, we note that here we are using spherical harmonics that are normalized differently to the ones in Section \ref{sec:Blm}.
The $Y_{\ell m}$ spherical harmonics in eq.~\eqref{eq:Ylm} are $4\pi$ normalized, while the ones we use in this section are Schmidt semi-normalized, $\mathcal{Y}_{\ell m} = Y_{\ell m}/\sqrt{2\ell+1}$, so that $\int \dd \Omega \, \mathcal{Y}_{\ell m} \, \mathcal{Y}_{\ell' m'}^* = 4\pi/(2\ell+1) \delta^K_{\ell \ell'} \delta^K_{m m'}$. 

The $L=0$ multipoles, $B_{\ell_1\ell_20}(k_1,k_2)$, describe the bispectrum monopole, while the $L > 0$ multipoles can only be non-zero in the presence of anisotropic RSD or AP effects.
We do not consider multipoles with odd $L$; while it is not immediately obvious that odd $L$ multipoles carry negligible information, our results will show that very little constraining power is lost by including only the even $L$ multipoles.
Since parity symmetry requires $\ell_1 + \ell_2 + L = $ even, if $L$ is even it follows that $\ell_1 + \ell_2$ is also even.
For these combinations of $(\ell_1, \ell_2, L)$, the $B_{\ell_1\ell_2L}$ multipoles are real-valued.

In this work, we make the global plane-parallel approximation so that the line-of-sight direction is fixed and determines the $\hat{z}$-axis for the spherical harmonic functions.
Since $\y_{LM}(\hat{z})=\delta^K_{M0}$, this simplifies the expressions for the TriPoSH multipoles in eqs.~\eqref{eq:Bl1l2L} and \eqref{eq:intYYYB} to
\begin{align}
	B_{\ell_1 \ell_2 L}(k_1,k_2) &= H_{\ell_1\ell_2L} \sum_{m} \wtj{\ell_1}{\ell_2}{L}{m}{-m}{0} 
	B_{\ell_1 \ell_2 L}^{m -m 0}(k_1,k_2) 
	\label{eq:Bl1l2L_global} \\
	B_{\ell_1 \ell_2 L}^{m -m 0}(k_1,k_2) &= N_{\ell_1 \ell_2 L}
	\int \frac{\diff^2 \hat{k}_1}{4\pi}
	\int \frac{\diff^2 \hat{k}_2}{4\pi}
	\y_{\ell_1}^{m*}(\hat{k}_1) \;
	\y_{\ell_2}^{-m*}(\hat{k}_2) \;
	B(\bk_1,\bk_2) \mathbb{1}(k_3).
\end{align}
In the last equation, we have added the factor $\mathbb{1}(k_3)$ to the integrand.
It is defined such that $\mathbb{1}(k_3) = 1$ if $k_3$ falls into the allowed $(k_\mathrm{min},k_\mathrm{max})$ range and zero otherwise.
In \citet{Sugiyama:2018yzo}, the integral above is over all triangles where two of the legs are $(k_1,k_2)$. Therefore, if  $k_1$ and $k_2$ can go as high as $k_\mathrm{max}$, then the allowed range of $k_3$ will go up to $2 \, k_\mathrm{max}$. 
However, this range of allowed triangles is different to the range of triangles that were used in the previous Fisher forecasts for the full bispectrum and $B_{\ell m}$ in Section \ref{sec:Blm}, where each leg of the triangle had to be in the same $k$ range, $k_\mathrm{min} \leq k_1, k_2, k_3 \leq k_\mathrm{max}$.
To make the forecasts cover the same $k$ range, we modify the integrals over $\hat{k}_1$ and $\hat{k}_2$ to impose the same $(k_\mathrm{min},k_\mathrm{max})$ limits on $k_3$:
\begin{align}
	\int \frac{\diff^2 \hat{k}_1}{4\pi}
	\int \frac{\diff^2 \hat{k}_2}{4\pi}
	&\rightarrow
	\int \frac{\diff^2 \hat{k}_1}{4\pi}
	\int \frac{\diff^2 \hat{k}_2}{4\pi}
	\mathbb{1}(k_3).
\end{align}

\subsection{$\hat{B}_{\ell_1 \ell_2 L}$ estimator and covariance}

In the global plane-parallel limit, the estimator for $B_{\ell_1 \ell_2 L}$ takes the same form as eq.~\eqref{eq:Bl1l2L_global} where the estimator for $B_{\ell_1 \ell_2 L}^{m -m 0}(k_1,k_2)$ is
\begin{align}
	\hat{B}_{\ell_1\ell_2 L}^{m_1 m_2 0}(k_1,k_2) &= N_{\ell_1 \ell_2 L} 
	\int \frac{\diff^2 \hat{k}_1}{4\pi} \y_{\ell_1}^{m_1*}(\hat{k}_1)
	\int \frac{\diff^2 \hat{k}_2}{4\pi} \y_{\ell_2}^{m_2*}(\hat{k}_2)
	\int \diff^3 k_3
	\delta_D(\bk_{123})
	\hat{\mathcal{B}}(\bk_1,\bk_2,\bk_3) \mathbb{1}(k_3) \nn \\
	&=  \frac{N_{\ell_1 \ell_2 L}}{N_\mathrm{modes}(k_1,k_2)} 
	\int_{k_1} \frac{\diff^3 q_1}{(2\pi)^3} \y_{\ell_1}^{m_1*}(\hat{q}_1)
	\int_{k_2} \frac{\diff^3 q_2}{(2\pi)^3} \y_{\ell_2}^{m_2*}(\hat{q}_2)
	\int \frac{\diff^3 q_3}{(2\pi)^3} (2\pi)^3 \delta_D(\bq_{123})
	\hat{\mathcal{B}}(\bq_1,\bq_2,\bq_3) \mathbb{1}(q_3).
\end{align}
In the second line we have exchanged the integrals over $\hat{k}_1$ and $\hat{k}_2$ for integrals over $\bq_1$ and $\bq_2$ in 3-dimensional Fourier-space shells and defined
\begin{align}
	N_\mathrm{modes}(k_1,k_2) &\equiv 
	\int_{k_1} \frac{\diff^3 q_1}{(2\pi)^3}
	\int_{k_2} \frac{\diff^3 q_2}{(2\pi)^3}
	\mathbb{1}(q_3).
\end{align}

Then the covariance for the TriPoSH multipoles is
\begin{align}
	\langle \hat{B}_{\ell_1 \ell_2 L}(k_1,k_2) \hat{B}_{\ell_1' \ell_2' L'}(k_1',k_2') \rangle
	&= H_{\ell_1 \ell_2 L} H_{\ell_1' \ell_2' L'}
	\sum_{m} \wtj{\ell_1}{\ell_2}{L}{m}{-m}{0} 
	\sum_{m'} \wtj{\ell_1'}{\ell_2'}{L'}{m'}{-m'}{0}
	\langle \hat{B}_{\ell_1 \ell_2 L}^{m -m 0}(k_1,k_2)
	\hat{B}_{\ell_1' \ell_2' L'}^{m' -m'0}(k_1',k_2') \rangle \\
	\langle \hat{B}_{\ell_1\ell_2L}^{m_1m_20}(k_1,k_2) \hat{B}_{\ell_1'\ell_2'L'}^{m_1'm_2'0}(k_1',k_2')\rangle 
	&= \frac{N_{\ell_1 \ell_2 L}}{N_\mathrm{modes}(k_1,k_2)} 
	\frac{N_{\ell_1' \ell_2' L'}}{N_\mathrm{modes}(k_1',k_2')} \nn \\
	&\times 
	\int_{k_1} \frac{\diff^3 q_1}{(2\pi)^3} \y_{\ell_1}^{m_1*}(\hat{q}_1) 
	\int_{k_2} \frac{\diff^3 q_2}{(2\pi)^3} \y_{\ell_2}^{m_2*}(\hat{q}_2)
	\int_{k_1'} \frac{\diff^3 q_1'}{(2\pi)^3} \y_{\ell_1'}^{m_1'*}(\hat{q}_1') 
	\int_{k_2'} \frac{\diff^3 q_2'}{(2\pi)^3} \y_{\ell_2'}^{m_2'*}(\hat{q}_2') \nn \\
	&\times \int \frac{\diff^3 q_3}{(2\pi)^3} (2\pi)^3 \delta_D(\bq_{123})
	\int \frac{\diff^3 q_3'}{(2\pi)^3} (2\pi)^3 \delta_D(\bq_{123}') 
	\langle \hat{\mathcal{B}} \hat{\mathcal{B}} \rangle
	\mathbb{1}(q_3)
	\mathbb{1}(q_3').
	\label{eq:Bl1l2L_covpart}
\end{align}
We substitute the Gaussian limit expression from eq.~\eqref{eq:covBB} for $\langle \hat{\mathcal{B}} \hat{\mathcal{B}} \rangle$ to find that the right side of eq.~\eqref{eq:Bl1l2L_covpart} has up to six terms. 
We write this as
\begin{align}
	\langle \hat{B}_{\ell_1\ell_2L}^{m_1m_20}(k_1,k_2) \hat{B}_{\ell_1'\ell_2'L'}^{m_1'm_2'0}(k_1',k_2')\rangle
	&= \frac{1}{V}
	\frac{N_{\ell_1 \ell_2 L}}{N_\mathrm{modes}(k_1,k_2)}
	\frac{N_{\ell_1' \ell_2' L'}}{N_\mathrm{modes}(k_1',k_2')}
	\sum_{i=1}^6
	\int_{k_1} \frac{\diff^3 q_1}{(2\pi)^3} \y_{\ell_1}^{m_1*}(\hat{q}_1)
	\int_{k_2} \frac{\diff^3 q_2}{(2\pi)^3} \y_{\ell_2}^{m_2*}(\hat{q}_2) \nonumber \\
	&\times 
	\int \frac{\diff^3 q_3}{(2\pi)^3} (2\pi)^3 \delta_D(\bq_{123})
	P_g(\bq_1) P_g(\bq_2) P_g(\bq_3) \times \mathcal{I}_i(\bq_1,\bq_2,\bq_3),
	\label{eq:Bl1l2L_covpart2}
\end{align}
and show the integrand $\mathcal{I}_i$ for each of the six terms in Table \ref{tab:Bl1l2L terms}.
The terms are numbered by the order in which they appear in eq.~\eqref{eq:covBB}.

\begin{table}
\begin{center}
  \begin{tabular}{ccllll}
    \hline
    $\mathcal{I}_1$ & $=$ & $\delta^K_{k_1 k_1'} \delta^K_{k_2 k_2'}$ 
      & $\y_{\ell_1'}^{m_1'*}(\hat{q}_1)$ 
      & $\y_{\ell_2'}^{m_2'*}(\hat{q}_2)$ 
      & $\mathbb{1}(q_3)$ \\
    $\mathcal{I}_2$ & $=$ & $\delta^K_{k_1 k_1'}$ 
      & $\y_{\ell_1'}^{m_1'*}(\hat{q}_1)$ 
      & 
      & $\Pi_{k_2'}(q_3) \y_{\ell_2'}^{m_2'*}(\hat{q}_3)$ \\
    $\mathcal{I}_3$ & $=$ & $\delta^K_{k_1 k_2'}$ 
      & $\y_{\ell_2'}^{m_2'*}(\hat{q}_1)$ 
      & 
      & $\Pi_{k_1'}(q_3) \y_{\ell_1'}^{m_1'*}(\hat{q}_3)$ \\
    $\mathcal{I}_4$ & $=$ & $\delta^K_{k_1 k_2'} \delta^K_{k_2 k_1'}$ 
      & $\y_{\ell_2'}^{m_2'*}(\hat{q}_1)$ 
      & $\y_{\ell_1'}^{m_1'*}(\hat{q}_2)$ 
      & $\mathbb{1}(q_3)$ \\
    $\mathcal{I}_5$ & $=$ & $\delta^K_{k_2 k_1'}$ 
      & 
      & $\y_{\ell_1'}^{m_1'*}(\hat{q}_2)$ 
      & $\Pi_{k_2'}(q_3) \y_{\ell_2'}^{m_2'*}(\hat{q}_3)$ \\
    $\mathcal{I}_6$ & $=$ & $\delta^K_{k_2 k_2'}$ 
      & 
      & $\y_{\ell_2'}^{m_2'*}(\hat{q}_2)$ 
      & $\Pi_{k_1'}(q_3) \y_{\ell_1'}^{m_1'*}(\hat{q}_3)$ \\
  \hline
  \end{tabular}
  \caption{Integrands $\mathcal{I}_i$ for each of the six terms of $\langle \hat{B}_{\ell_1\ell_2L}^{m_1m_20}(k_1,k_2) \hat{B}_{\ell_1'\ell_2'L'}^{m_1'm_2'0}(k_1',k_2')\rangle$ in eq.~\eqref{eq:Bl1l2L_covpart2}.
  We have defined functions $\Pi_{k}(q)$, such that $\Pi_{k}(q)=1$ if $q$ falls into the $k$ bin and zero otherwise, and $\mathbb{1}(k_3)$, such that $\mathbb{1}(k_3) = 1$ if $k_3$ falls into the allowed $(k_\mathrm{min},k_\mathrm{max})$ range and zero otherwise.
  }
  \label{tab:Bl1l2L terms}
\end{center}
\end{table}

As we did for the covariance of the $B_{\ell 0}$ multipoles in Section \ref{sec:Blm}, after rewriting the Dirac delta function $\delta_D(\bq_{123})$ using eq.~\eqref{eq:dirac_delta}, we can compute the covariance for the TripoSH multipoles using 3D FFTs.
We found that this allowed for a relatively fast computation of the covariance matrix that also yielded an invertible covariance matrix.\footnote{We also used the thin-shell approximation to compute the covariance using SHTOOLS, not based on 3D FFTs, but this calculation was computationally very expensive and yielded a covariance matrix that experienced severe numerical issues in our implementation.}

\subsection{$B_{\ell_1\ell_2 L}$ Fisher forecast results}
\label{sec:Bl1l2L_fisher}

\begin{figure*}
\centering
\includegraphics[width=\textwidth]{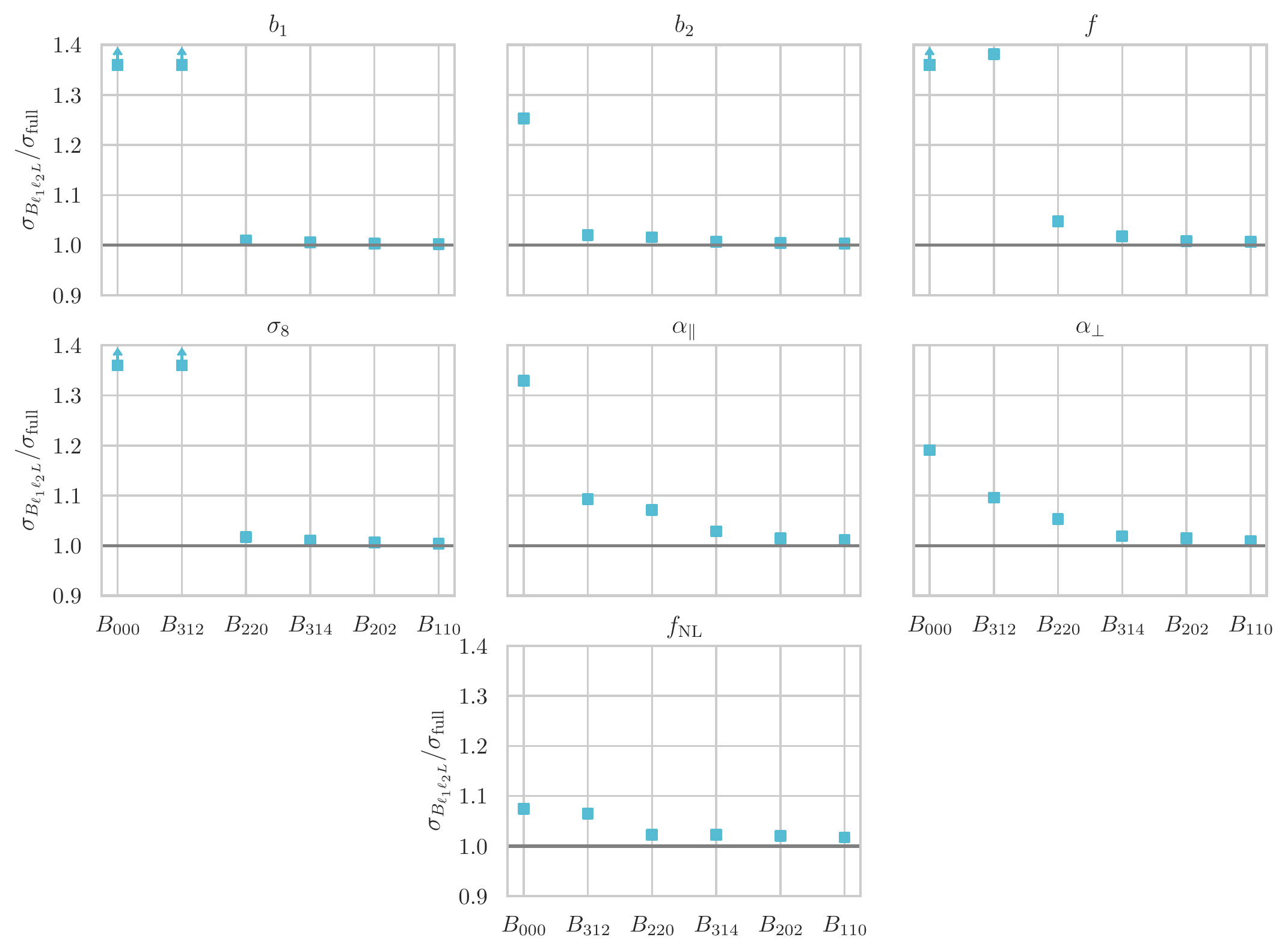}
\caption{Forecasted constraints from $B_{\ell_1 \ell_2 L}$ including an increasing number of multipoles compared to constraints from the full redshift-space bispectrum. A value of $\sigma_{B_{\ell_1 \ell_2 L}}/\sigma_\mathrm{full} = 1$ would indicate that the multipoles recover all of the available information on the parameters, $(b_1, b_2, f, \sigma_8, \alpha_\parallel, \alpha_\perp, \fnl)$.
In each panel, the leftmost point includes only $B_{000}$, the next point above $B_{312}$ includes $(B_{000},B_{312})$, and so on, such that the rightmost point includes six multipoles, $(B_{000},B_{312},B_{220},B_{314},B_{202},B_{110})$.
The first three multipoles are sufficient to obtain constraints that are within 8 per cent of the full bispectrum results, while including the six multipoles shown here will obtain constraints that are within 2 per cent of the full bispectrum.
}
\label{fig:Bl1l2L_fisher}
\end{figure*}

The Fisher matrix is
\begin{align}
	\mathbfss{F}_{ij} &\equiv \sum_{\ell_1 \ell_2 L} \, \sum_{k_1 k_2} \,
	\sum_{\ell_1' \ell_2' L'} \, \sum_{k_1' k_2'}
	\frac{\partial B_{\ell_1 \ell_2 L}(k_1,k_2)}{\partial \theta_i} 
	\mathbfss{C}^{-1}[B_{\ell_1 \ell_2 L}(k_1,k_2), B_{\ell_1' \ell_2' L'}(k_1',k_2')]
	\frac{\partial B_{\ell_1' \ell_2' L'}(k_1',k_2')}{\partial \theta_j}
	+\mathbfss{F}^P_{ij}.
\end{align}

We include all even $L$ up to $L_\mathrm{max}=6$. For each $L$, in principle there are an unlimited number of $(\ell_1,\ell_2)$ pairs that contribute to the bispectrum signal, that satisfy $\ell_1 + \ell_2 + L = $ even and the triangle inequality. In this work we consider all allowed pairs of $(\ell_1,\ell_2)$ with $\ell_1, \ell_2 \leq \ell_\mathrm{max}=6$. This results in a total of 40 possible $(\ell_1,\ell_2,L)$ combinations.

For $k_1$ and $k_2$, we use the same $k$-binning as in Section \ref{sec:Blm} for the $B_{\ell m}$ multipoles: $k_\mathrm{min} = 0.02 \, h \, \Mpc^{-1}$, $k_\mathrm{max} = 0.15 \, h \, \Mpc^{-1}$, $\Delta k = 0.01 \, h \, \Mpc^{-1}$. For a multipole with $\ell_1 \neq \ell_2$, there are 196 $(k_1,k_2)$ bins. When $\ell_1 = \ell_2$, there are 105 $(k_1,k_2)$ bins. 

We take the most minimal data set to be $B_{000}$, but after that it is not obvious which multipoles to include for the most efficient multipole compression. Since we wish to find the most efficient compression, we aim to include only as many multipoles as is necessary, and not all them, as this would amount to a data vector that is 5,838 bins in length.
Instead, we construct the list of sufficient multipoles as follows. 
After $B_{000}$, we choose the next multipole to be whichever one out of the remaining 39 multipoles gives the largest improvements in the forecasted constraints. 
For a given starting set of multipoles $X$, we define an improvement factor for each prospective multipole $Y$ as
\begin{equation}
	\prod_{i=1}^7 \frac{\sigma_{\theta_i}(X)}{\sigma_{\theta_i}(X \cup Y)},
	\label{eq:improvement_fac}
\end{equation}
and we choose $Y$ such that this factor is maximized.
In this case, when $X = \{ B_{000} \}$, we computed this factor for each of the 39 other remaining multipoles and found that it was largest when $Y = B_{312}$.
In the next iteration, $X = \{ B_{000}, B_{312} \}$ and then eq.~\eqref{eq:improvement_fac} is maximised when $Y = B_{220}$.
We repeat this process until the constraints do not show a significant change with the addition of any further multipoles.
For each set of multipoles, we compare the resulting forecast to constraints from the full $z$-space bispectrum (combined with the power spectrum) to obtain the results shown in  Fig.~\ref{fig:Bl1l2L_fisher}.
We find that three multipoles, $(B_{000}, B_{312}, B_{220})$, (406 bins in total) are sufficient to recover constraints that are within 8 per cent of the full bispectrum result. 
Beyond this, by adding three more multipoles, $(B_{314},B_{202},B_{110})$, (903 bins in total), constraints are within 2 per cent of the full bispectrum result.


\section{Modal basis decomposition}
\label{sec:modal}

Finally, in this section we discuss the modal approach to compressing the redshift-space bispectrum.
The modal decomposition of the bispectrum is motivated by the fact that the galaxy bispectrum should be a relatively smooth function of the triangle and its orientation.
This implies that for a well-chosen set of basis functions, the modal decomposition should converge efficiently with few modes.

Previous work analysing mock halo catalogs has shown that the modal decomposition works very well to recover constraints on galaxy bias and shot noise parameters as the standard bispectrum estimator in real space, without RSD \citep{Byun:2020rgl}.
This work is an extension of the modal decomposition method to redshift space, demonstrating the performance for the tree-level redshift-space bispectrum in a Fisher forecast.
We note that the modal decomposition method is general--at its core it is simply using a separable basis to expand the bispectrum--, and what we implement in this work is one specific form that this basis may take.
There are other options for basis choices that could be explored and compared, such as the modal decomposition in \citet{Regan:2017vgi} that is based on Legendre multipoles.

\subsection{Modal decomposition}
\label{sec:modal_intro}

In the modal decomposition approach, the weighted bispectrum is approximated as a linear combination of basis functions, 
\begin{equation}
	wB_g(\bk_1,\bk_2,\bk_3) = \sum_{n=0}^{N_\mathrm{modes}} \beta^Q_n Q_n(\bk_1,\bk_2,\bk_3),
\end{equation}
where $w$ is a weighting function that in general depends on the three $\bk_i$ wave-vectors, 
$\beta^Q_n$ are the modal expansion coefficients, and $Q_n$ are the basis functions.
If the basis functions are chosen such that only relatively small $N_\mathrm{modes}$ is necessary to reconstruct $B$ to sufficient accuracy, then the modal compression would be very efficient.

To solve for the modal coefficients given a model for $B$, we first define the inner product, 
\begin{equation}
	\llangle X | Y \rrangle \equiv 
	\int_{\bk_1}
	\int_{\bk_2}
	\int_{\bk_3}
	(2\pi)^3 \delta_D(\bk_{123})  
	\frac{X(\bk_1,\bk_2,\bk_3) Y(\bk_1,\bk_2,\bk_3)}{k_1 k_2 k_3},
\end{equation}
which integrates the product of two functions over all triangle shapes and orientations.
Then we can solve the linear equation 
\begin{equation}
	\llangle wB | Q \rrangle = \vec{\beta}^Q \cdot \gamma 
	\label{eq:modal_lineq}
\end{equation}
to obtain the $\beta^Q_n$, where we have defined the positive-definite symmetric matrix $\gamma \equiv \llangle Q | Q \rrangle$.

This modal decomposition method so far applies generally in real space and redshift space. 
Compared to previous work in real space (e.g. \citet{Byun:2020rgl}), the optimal weighting in redshift space is different,
\begin{equation}
	w \equiv \frac{\sqrt{k_1 k_2 k_3}}{\sqrt{P_g(\bk_1) P_g(\bk_2) P_g(\bk_3)}},
\end{equation}
and it will be necessary to choose our $Q_n$ basis functions to have an angular dependence so that the the anisotropic component of the bispectrum, induced by RSD and AP effects, can be captured efficiently by the modal expansion.

\subsection{Custom modes}
\label{subsec:custom_modes}

While the $Q_n$ basis functions can in principle be constructed from any set of basis functions, in previous work on the real-space modal bispectrum, it was advantageous to build a basis of separable \textit{custom modes} that were functions that could analytically reproduce the tree-level bispectrum model exactly \citep{Hung:2019ygc,Byun:2020rgl}. 
We apply the same approach here to the redshift-space bispectrum.

The tree-level redshift-space bispectrum expression in eq.~\eqref{eq:Btree} can be rewritten exactly as a sum of 83 separable $Q_n$ basis functions,\footnote{We note that if we fix $\fnl=0$ in the modeling, then only 38 $Q_n$ are necessary to recover eq.~\eqref{eq:Btree}.} where each $Q_n$ is made up of a combination of three $q_n(k,\mu)$ functions from Table \ref{tab:qn},
\begin{equation}
	Q_n(k_1,k_2,k_3,\mu_1,\mu_2,\mu_3) = q_{\{p}(k_1,\mu_1) q_r(k_2,\mu_2)q_{s\}}(k_3,\mu_3).
\end{equation}
The curly brackets around the $p$, $r$, and $s$ subscripts signify that the function is symmetrized over permutations of $p$, $r$ and $s$.
We refer to this basis of $Q_n$ functions as \textit{custom modes}.
For compactness, we will sort basis functions based on increasing values of $p+r+s$. 
Then the first five $(p,r,s)$ combinations are $(0,0,0)$, $(0,0,1)$, $(0,1,1)$, $(0,2,3)$, $(0,0,7)$ and the last five combinations are $(16,16,23)$, $(16,19,22)$, $(16,19,23)$, $(19,19,22)$, $(19,19,23)$.

In the case where the AP parameters are fixed to unity, $\alpha_\parallel = \alpha_\perp = 1$, it is straightforward to derive the corresponding coefficients $\beta^Q_n$ for this basis, given the parameters of interest in this work: $b_1$, $b_2$, $f$, $\sigma_8$, $\alpha_\parallel$, $\alpha_\perp$, and $\fnl$.
However, when the AP parameters are varied, the shape of the bispectrum changes in a way that may or may not be well-described by the existing set of 83 custom modes.
Hence we also consider a larger basis of \textit{extended custom modes} which includes additional modes defined to capture the changes to the custom $Q_n$ induced by the AP parameters.
Therefore, the extended custom basis consists of 83 custom modes, 82 functions for $\partial_{\alpha_\parallel}Q_n$, and 82 functions for $\partial_{\alpha_\parallel}Q_n$, for a total of 247 basis functions. 
There are only 82 derivative functions for each AP parameter, because one of the custom $Q_n$ is a constant (when $p=r=s=0$), so its derivative with respect to the AP parameters is zero.

For each choice of basis, the custom modes and extended custom modes, we use FFTs to calculate the $\gamma$ matrix consisting of inner products between all pairs of $Q_n$, in the same manner as in \citet{Byun:2020rgl}.

\begin{table*}
\begin{center}
  \begin{tabular}{r|c||r|c||r|c||r|c}
    $n$ & $q_n(k,\mu)$ & $n$ & $q_n(k,\mu)$ & $n$ & $q_n(k,\mu)$ & $n$ & $q_n(k,\mu)$ \\
    \hline
    0 & 1 & 6 & $\mu^2$ & 12 & $P_m(k) \mu / k$ & 18 & $P_m(k)/k^2\mathcal{M}(k)$\\
    1 & $P_m(k)$ & 7 & $P_m(k) \mu^2$ & 13 & $P_m(k) \mu^3 / k$ & 19 & $P_m(k)\mu^2/\mathcal{M}(k)$ \\
    2 & $P_m(k) k^2$ & 8 & $P_m(k) \mu^2 k^2$ & 14 & $P_m(k) \mu^4$ & 20 & $P_m(k)\mu^2k^2/\mathcal{M}(k)$\\
    3 & $P_m(k) / k^2$ & 9 & $P_m(k) \mu^2 / k^2$ & 15 & $\mu k$ & 21 & $P_m(k)\mu/k\mathcal{M}(k)$ \\
    4 & $k^2$ & 10 & $\mu^2 k^2$ & 16 & $P_m(k)/\mathcal{M}(k)$ & 22 & $\mathcal{M}(k)$ \\
    5 & $k^4$ & 11 & $\mu^2 k^4$ & 17 & $P_m(k)k^2/\mathcal{M}(k)$ & 23 & $\mu^2\mathcal{M}(k)$
  \end{tabular}
  \caption{Set of 24 $q_n(k,\mu)$ functions for recovering the tree-level bispectrum. $P_m(k)$ is the linear matter power spectrum. $\mathcal{M}(k)$ depends on the matter transfer function and is defined in Section \ref{subsec:modeling}.}
  \label{tab:qn}
\end{center}
\end{table*}

\subsection{Orthonormal modes and covariance}
\label{sec:modal_ortho_modes}

The advantage of the $Q_n$ basis is that given input parameters, $(b_1,b_2,f,\sigma_8,\alpha_\parallel,\alpha_\perp,\fnl)$, we can analytically predict the $\beta^Q_n$ expansion coefficients.
On the other hand, it is also convenient to rotate to another basis where the basis functions are orthogonal, so that the expansion coefficients on this basis are uncorrelated.
We will use a set of orthonormal basis functions called $R_n$ which is defined by having $\llangle R_n|R_m \rrangle = \delta^K_{nm}$.
On this basis then, 
\begin{equation}
	\beta^R_n = \llangle wB | R_n \rrangle,
\end{equation}
and to rotate expansion coefficients from the $\beta^Q$ to the $\beta^R$ is
\begin{equation}
	\beta^R = \lambda^T \beta^Q,
\end{equation}
where $\lambda$ is the lower-triangular matrix that results from the Cholesky decomposition of $\gamma$, $\gamma = \lambda \lambda^T$.

\subsection{Modal estimator and covariance}
\label{sec:modal_est_and_cov}

From eq.~\eqref{eq:modal_lineq}, the estimator for the orthonormal modal coefficients is
\begin{equation}
	\hat{\beta}^R_n = \llangle w\mathcal{B} | R_n \rrangle,
\end{equation}
and the covariance is 
\begin{align}
	\langle \hat{\beta}^R_n \hat{\beta}^R_m \rangle
	&=
	\int \frac{\diff^3 k_1}{(2\pi)^3}
	\int \frac{\diff^3 k_2}{(2\pi)^3}	
	\int \frac{\diff^3 k_3}{(2\pi)^3}
	(2\pi)^3 \delta_D(\bk_{123})  
	\frac{ w R_n(\bk_1,\bk_2,\bk_3)}{k_1 k_2 k_3} \nonumber  \\
	&\times 
	\int \frac{\diff^3 k_1'}{(2\pi)^3}
	\int \frac{\diff^3 k_2'}{(2\pi)^3}	
	\int \frac{\diff^3 k_3'}{(2\pi)^3}
	(2\pi)^3 \delta_D(\bk_{123}')
	\frac{w R_m(\bk_1',\bk_2',\bk_3')}{k_1' k_2' k_3'}
	\llangle
	\mathcal{B}(\bk_1,\bk_2,\bk_3)\mathcal{B}(\bk_1',\bk_2',\bk_3')
	\rrangle.
\end{align}
Using eq.~\eqref{eq:covBB} for $\llangle \mathcal{B} \mathcal{B}' \rrangle$, the Gaussian covariance matrix for $\beta^R_n$ simplifies to
\begin{align}
	\langle \hat{\beta}^R_n \hat{\beta}^R_m \rangle
	&= \frac{6}{V} \delta^K_{nm}.
\end{align}

\subsection{Note on computing derivatives with respect to AP parameters}
\label{sec:modal_derivatives}

Unlike the other parameters $(b_1,b_2,f,\sigma_8,\fnl)$, the effect of the AP parameters $(\alpha_\parallel,\alpha_\perp)$ on the the custom $\beta^Q$ coefficients is not trivial to write down analytically because the AP parameters distort the $k$-dependent shape of the bispectrum.
For general values of the AP parameters, we can still compute the tree-level bispectrum \textit{exactly} as
\begin{equation}
	\tilde{B}_\mathrm{tree}(\bk,\bk_2,\bk_3,\alpha) = \sum_{n=0}^{82} \tilde{\beta}^Q_n \tilde{Q}_n(\bk_1,\bk_2,\bk_3,\alpha).
	\label{eq:Btilde}
\end{equation}
Here, each $\tilde{Q}_n$ has the same functional form as the custom modes, $Q_n$, but the arguments are now wave-vectors that are rescaled by the AP parameters. 
Correspondingly, $\tilde{\beta}^Q_n$ have the same dependence on the input parameters $(b_1,b_2,f,\sigma_8,\fnl)$ as before, but now they are additionally divided by the appropriate AP factors, $\tilde{\beta}^Q_n = \beta^Q_n/(\alpha_\parallel \alpha_\perp^2)^2$.

We would like to find the modal coefficients $\beta^Q$ that can reconstruct this $\tilde{B}_\mathrm{tree}$,
\begin{equation}
	w(\bk) \tilde{B}_\mathrm{tree}(\bk,\alpha) = \sum_m \beta^Q_m Q_m(\bk).
	\label{eq:wBtilde}
\end{equation}
Combining eqs.~\eqref{eq:Btilde} and \eqref{eq:wBtilde}, we find $\beta^Q$ by solving the linear equation
\begin{equation}
	\gamma \cdot \beta^Q = \llangle Q |  w\tilde{Q} \rrangle \cdot \tilde{\beta}^Q.
\end{equation}

While the custom modes can reconstruct the fiducial bispectrum with $\alpha_\parallel = \alpha_\perp = 1$ exactly, we do not know \textit{a priori} whether they are efficient at capturing changes to the bispectrum induced by varying the AP parameters.

\subsection{Fisher forecast results}
\label{sec:modal_fisher}

\begin{figure*}
\centering
\includegraphics[width=\textwidth]{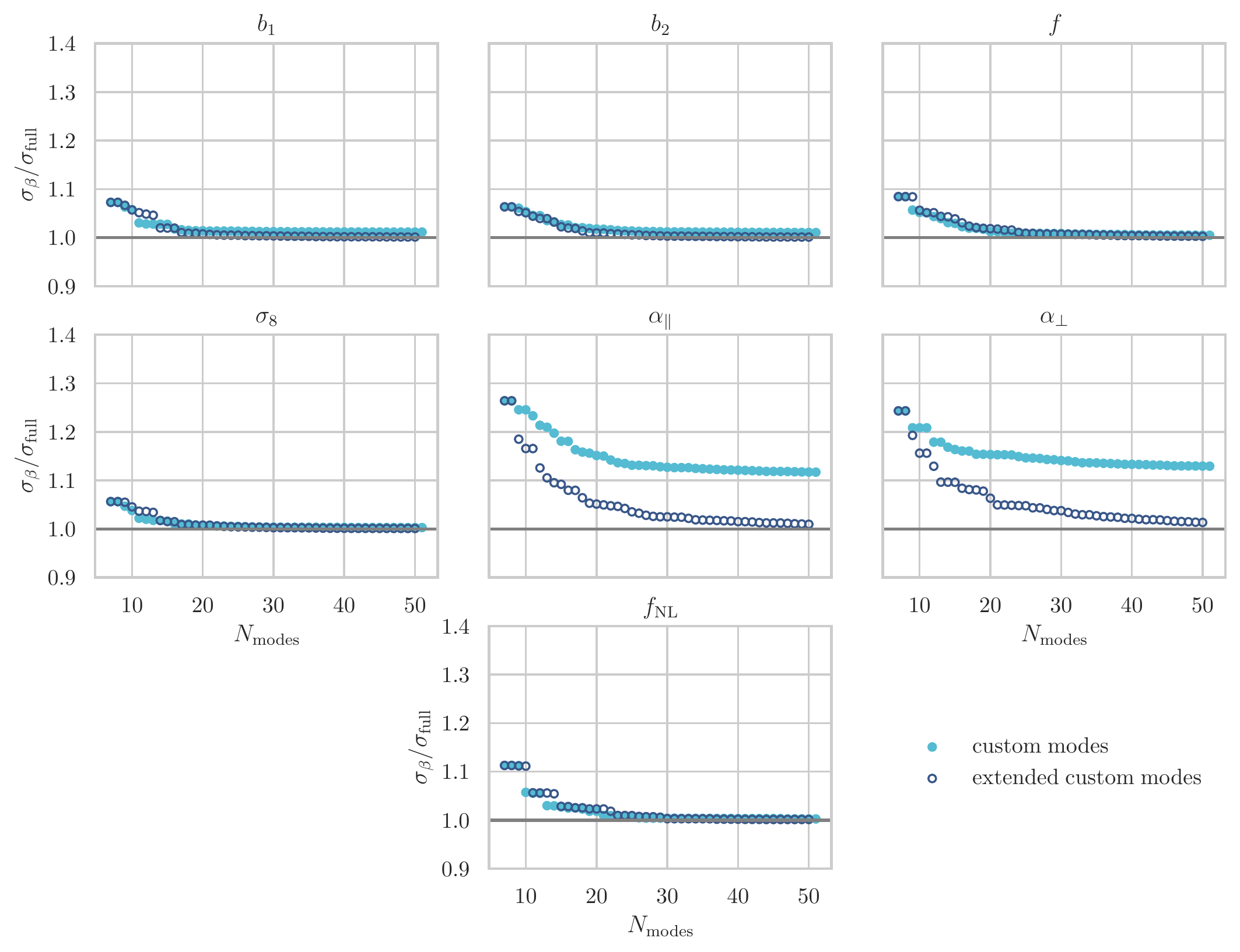}
\caption{Forecasted constraints from the modal decomposition of the bispectrum including an increasing number of basis functions, $N_\mathrm{modes}$, compared to constraints from the full redshift-space bispectrum. 
The basis of custom modes (filled light blue circles) are not sufficient to recover constraints on the AP parameters to within 10 per cent of the full bispectrum constraint.
However, the basis of extended custom modes (empty dark blue circles) requires only 14 modes to recover constraints to within 10 per cent or 42 modes to recover constraints to within 2 per cent, of the full bispectrum.}
\label{fig:modal}
\end{figure*}

The Fisher matrix is
\begin{align}
	\mathbfss{F}_{ij} &\equiv \sum_{nm} 
	\frac{\partial \beta^R_n}{\partial \theta_i} 
	\mathbfss{C}^{-1}[\beta^R_n,\beta^R_m]
	\frac{\partial \beta^R_m}{\partial \theta_j} 
	+\mathbfss{F}^P_{ij} \nonumber \\
	&= \frac{V}{6} \sum_n 
	\frac{\partial \beta^R_n}{\partial \theta_i} 
	\frac{\partial \beta^R_n}{\partial \theta_j}
	+ \mathbfss{F}^P_{ij}.
\end{align}

The comparison of constraints from the modal bispectrum vs the full redshift-space bispectrum is shown in Fig.~\ref{fig:modal} as a function of the number of modes included. 
We show the results for both the custom modes and the extended custom modes.
To show the most efficient compression, we order the modes in the same way that we ordered the $B_{\ell_1\ell_2L}$ multipoles in Section \ref{sec:Bl1l2L_fisher}. 
The most minimal set of modes requires at least 7, since we are forecasting constraints for 7 parameters, and we choose these starting modes to be the 7 modes with the smallest $p+r+s$ values.

We find that using only the original set of 83 custom modes (light blue circles in Fig.~\ref{fig:modal}), the modal constraints do not converge to within 10 per cent of the constraints from the full bispectrum.
This is because the AP parameters change the bispectrum's dependence on triangles and orientations in a way that is not captured by the custom modes, but this issue is resolved when the extended custom modes are used (dark blue circles).
In this case, only 14 modes are necessary to recover constraints that are within 10 per cent of the full bispectrum result, and only 42 modes are required to obtain constraints that are within 2 per cent. 
Within the set of the $N_\mathrm{modes}= 14$ modes, the fact that two of these are from the extended custom basis (and not included in the original 83 custom modes) is what allows the constraints from the extended basis to recover constraints that are closer to what is obtained from the full bispectrum.
Similarly, within the set of $N_\mathrm{modes}= 42$ modes, 10 of these are only in the extended basis.

We emphasise that the extended custom modes are only necessary if we are interested in constraining the AP factors.
Performing the same forecast as above, except including only the parameters $(b_1,b_2,f,\sigma_8,\fnl)$, we find that a basis of only 13 of the original (non-extended) custom modes is sufficient to obtain forecasted errors that are within 2 per cent of the full bispectrum result.


\section{Discussion}
\label{sec:discussion}

\begin{figure}
\centering
\includegraphics[width=0.6\textwidth]{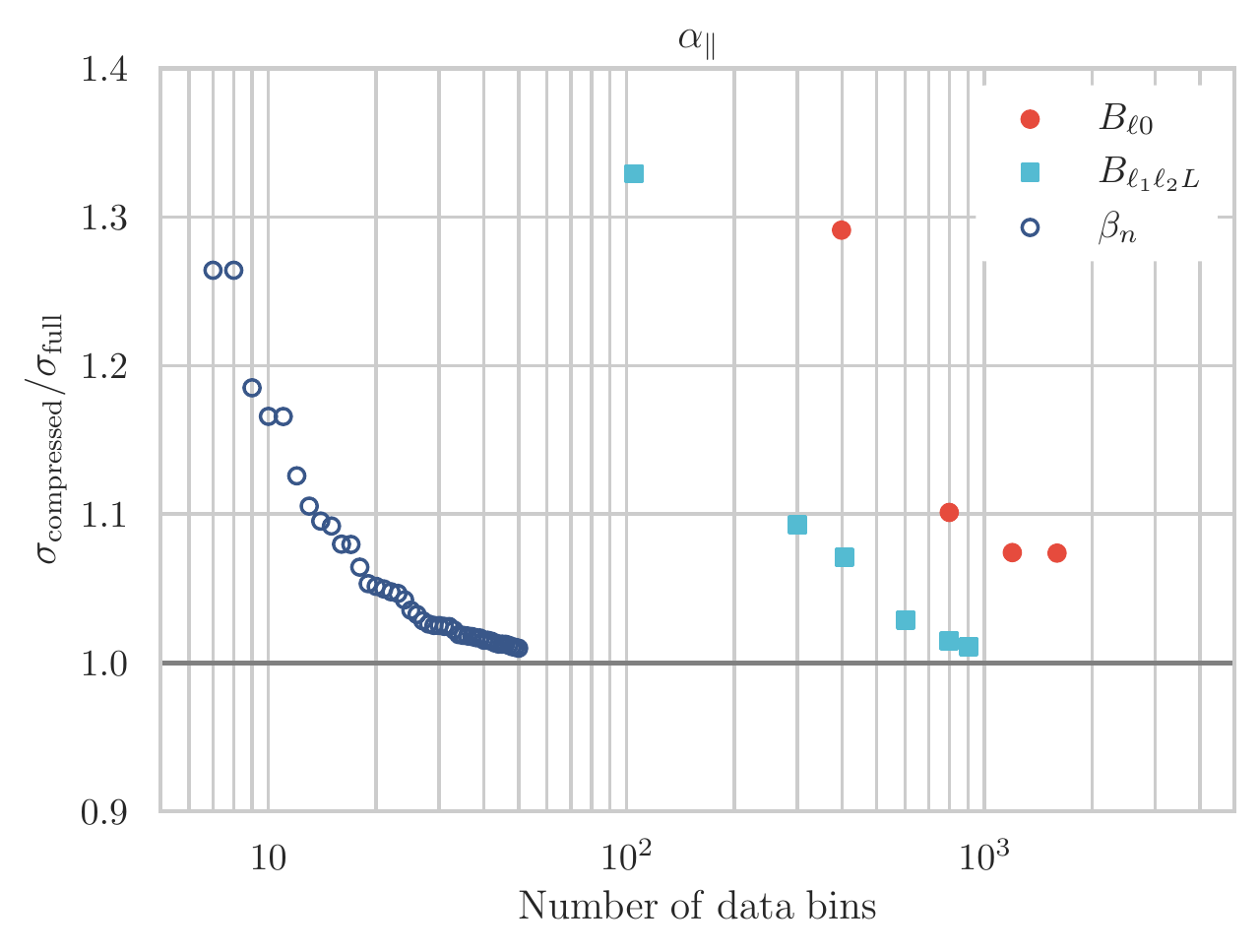}
\caption{Per cent difference between the forecasted constraint on the AP parameter, $\alpha_\parallel$, from the $B_{\ell 0}$ multipoles (red circles), TripoSH $B_{\ell_1 \ell_2 L}$ multipoles (light blue squares), and modal coefficients $\beta_n$ (empty dark blue circles) compared to the forecasted constraint from the full redshift-space bispectrum, $\sigma_\mathrm{full}$.
The data shown here are the same as the data shown in the $\alpha_\parallel$ panels of Figs.~\ref{fig:Bl0_fisher_fft}, \ref{fig:Bl1l2L_fisher}, and \ref{fig:modal}, except here the $x$-axis is the number of data bins that are used in the forecast to obtain $\sigma_\mathrm{compressed}$.
While all three compression schemes can recover constraints that are within 10 per cent of the full bispectrum constraint, this figure shows that the modal decomposition requires far fewer data bins.
}
\label{fig:comparison}
\end{figure}

In this work, we extended the modal decomposition method to the redshift-space bispectrum and compared its performance and efficiency with two multipole decompositions in the literature: the spherical harmonic $B_{\ell m}(k_1,k_2,k_3)$ multipoles \citep{Scoccimarro:2015bla} and the tripolar spherical harmonic $B_{\ell_1 \ell_2 L}(k_1,k_2)$ multipoles \citep{Sugiyama:2018yzo}.
We compared these three bispectrum compression schemes by computing Fisher forecasted constraints on a basic set of cosmological and galaxy bias parameters, $(b_1, b_2, f, \sigma_8, \alpha_\parallel, \alpha_\perp, \fnl)$, for a reference survey.
In each case, we compared the constraint from the compressed statistic to the constraint from the full redshift-space bispectrum.
We find that the modal bispectrum can recover the same forecasted constraining power as the full bispectrum to within 10 (2) per cent by using 14 (42) modal expansion coefficients. 
A comparable level of agreement can also be achieved in our forecasts for the other multipole decompositions, $B_{\ell m}$ and $B_{\ell_1 \ell_2 L}$, through using a much larger data vector.
This is illustrated for one of the parameters, $\alpha_\parallel$, in Fig.~\ref{fig:comparison}.
We have chosen to show the comparison between estimators for this particular parameter, because this is the parameter that typically required more multipoles or modes for constraints to converge.

Further development of the modal decomposition approach for the galaxy bispectrum could take different directions.
As this work is based on a tree-level bispectrum model, it remains to be tested with simulations or mock catalogs how well the bases considered here could reconstruct the non-linear matter or galaxy bispectrum.
Furthermore, as current and future galaxy surveys observe larger areas of the sky, it will become necessary to move beyond bispectrum estimators based on the global plane parallel approximation, considered in this work.
This has been explored for the $B_{\ell m}$ and $B_{\ell_1 \ell_2 L}$ multipoles in \citet{Scoccimarro:2015bla} and \citet{Sugiyama:2018yzo}, but we leave a corresponding modal basis to future work.
Another layer of realism that needs to be included in future is how the survey window function is incorporated in the modeling of the modal coefficients. 
Recent work on windowless bispectrum estimation \citep{Philcox:2021ukg} and window function convolutions \citet{Pardede:2022udo} may be transferable to the modal approach as well.


\section*{Acknowledgements}

JB would like to thank Chen Heinrich for helpful discussions and code comparisons.
We acknowledge support from the SPHEREx project under a contract from the NASA/GODDARD Space Flight Center to the California Institute of Technology. 
This research was also supported by the David and Lucile Packard Foundation.
Part of this work is based upon High Performance Computing (HPC) resources supported by the University of Arizona TRIF, UITS, and Research, Innovation, and Impact (RII) and maintained by the UArizona Research Technologies department.


\section*{Data Availability}

The data underlying this article will be shared on reasonable request to the corresponding author.



\bibliographystyle{mnras}
\bibliography{references}

\begin{thebibliography}{}
\makeatletter
\relax
\def\mn@urlcharsother{\let\do\@makeother \do\$\do\&\do\#\do\^\do\_\do\%\do\~}
\def\mn@doi{\begingroup\mn@urlcharsother \@ifnextchar [ {\mn@doi@}
  {\mn@doi@[]}}
\def\mn@doi@[#1]#2{\def\@tempa{#1}\ifx\@tempa\@empty \href
  {http://dx.doi.org/#2} {doi:#2}\else \href {http://dx.doi.org/#2} {#1}\fi
  \endgroup}
\def\mn@eprint#1#2{\mn@eprint@#1:#2::\@nil}
\def\mn@eprint@arXiv#1{\href {http://arxiv.org/abs/#1} {{\tt arXiv:#1}}}
\def\mn@eprint@dblp#1{\href {http://dblp.uni-trier.de/rec/bibtex/#1.xml}
  {dblp:#1}}
\def\mn@eprint@#1:#2:#3:#4\@nil{\def\@tempa {#1}\def\@tempb {#2}\def\@tempc
  {#3}\ifx \@tempc \@empty \let \@tempc \@tempb \let \@tempb \@tempa \fi \ifx
  \@tempb \@empty \def\@tempb {arXiv}\fi \@ifundefined
  {mn@eprint@\@tempb}{\@tempb:\@tempc}{\expandafter \expandafter \csname
  mn@eprint@\@tempb\endcsname \expandafter{\@tempc}}}

\bibitem[\protect\citeauthoryear{Ade et~al.}{Ade et~al.}{2014}]{Ade:2013ydc}
Ade P.,  et~al., 2014, \mn@doi [Astron. Astrophys.]
  {10.1051/0004-6361/201321554}, 571, A24

\bibitem[\protect\citeauthoryear{Ade et~al.}{Ade et~al.}{2016}]{Planck:2015igc}
Ade P. A.~R.,  et~al., 2016, \mn@doi [Astron. Astrophys.]
  {10.1051/0004-6361/201526681}, 594, A16

\bibitem[\protect\citeauthoryear{Agarwal, Desjacques, Jeong  \&
  Schmidt}{Agarwal et~al.}{2021}]{Agarwal:2020lov}
Agarwal N.,  Desjacques V.,  Jeong D.,   Schmidt F.,  2021, \mn@doi [JCAP]
  {10.1088/1475-7516/2021/03/021}, 03, 021

\bibitem[\protect\citeauthoryear{Aghamousa et~al.}{Aghamousa
  et~al.}{2016}]{DESI:2016fyo}
Aghamousa A.,  et~al., 2016

\bibitem[\protect\citeauthoryear{Aghanim et~al.}{Aghanim
  et~al.}{2020}]{Planck:2018vyg}
Aghanim N.,  et~al., 2020, \mn@doi [Astron. Astrophys.]
  {10.1051/0004-6361/201833910}, 641, A6

\bibitem[\protect\citeauthoryear{Alcock \& Paczynski}{Alcock \&
  Paczynski}{1979}]{Alcock:1979mp}
Alcock C.,  Paczynski B.,  1979, \mn@doi [Nature] {10.1038/281358a0}, 281, 358

\bibitem[\protect\citeauthoryear{Alsing \& Wandelt}{Alsing \&
  Wandelt}{2018}]{Alsing:2017var}
Alsing J.,  Wandelt B.,  2018, \mn@doi [Mon. Not. Roy. Astron. Soc.]
  {10.1093/mnrasl/sly029}, 476, L60

\bibitem[\protect\citeauthoryear{Baldauf, Seljak, Desjacques  \&
  McDonald}{Baldauf et~al.}{2012}]{Baldauf:2012hs}
Baldauf T.,  Seljak U.,  Desjacques V.,   McDonald P.,  2012, \mn@doi [Phys.
  Rev. D] {10.1103/PhysRevD.86.083540}, 86, 083540

\bibitem[\protect\citeauthoryear{Barreira}{Barreira}{2021}]{Barreira:2021ueb}
Barreira A.,  2021

\bibitem[\protect\citeauthoryear{Bernardeau, Colombi, Gaztanaga  \&
  Scoccimarro}{Bernardeau et~al.}{2002}]{Bernardeau:2001qr}
Bernardeau F.,  Colombi S.,  Gaztanaga E.,   Scoccimarro R.,  2002, \mn@doi
  [Phys. Rept.] {10.1016/S0370-1573(02)00135-7}, 367, 1

\bibitem[\protect\citeauthoryear{Blanchard et~al.}{Blanchard
  et~al.}{2020}]{Euclid:2019clj}
Blanchard A.,  et~al., 2020, \mn@doi [Astron. Astrophys.]
  {10.1051/0004-6361/202038071}, 642, A191

\bibitem[\protect\citeauthoryear{Blas, Lesgourgues  \& Tram}{Blas
  et~al.}{2011}]{Blas:2011rf}
Blas D.,  Lesgourgues J.,   Tram T.,  2011, \mn@doi [JCAP]
  {10.1088/1475-7516/2011/07/034}, 07, 034

\bibitem[\protect\citeauthoryear{Bose \& Taruya}{Bose \&
  Taruya}{2018}]{Bose:2018zpk}
Bose B.,  Taruya A.,  2018, \mn@doi [JCAP] {10.1088/1475-7516/2018/10/019}, 10,
  019

\bibitem[\protect\citeauthoryear{Bose, Byun, Lacasa, Moradinezhad~Dizgah  \&
  Lombriser}{Bose et~al.}{2020}]{Bose:2019wuz}
Bose B.,  Byun J.,  Lacasa F.,  Moradinezhad~Dizgah A.,   Lombriser L.,  2020,
  \mn@doi [JCAP] {10.1088/1475-7516/2020/02/025}, 02, 025

\bibitem[\protect\citeauthoryear{Byun, Eggemeier, Regan, Seery  \& Smith}{Byun
  et~al.}{2017}]{Byun:2017fkz}
Byun J.,  Eggemeier A.,  Regan D.,  Seery D.,   Smith R.~E.,  2017, \mn@doi
  [Mon. Not. Roy. Astron. Soc.] {10.1093/mnras/stx1681}, 471, 1581

\bibitem[\protect\citeauthoryear{Byun, Franco, Howlett, Bonvin  \&
  Obreschkow}{Byun et~al.}{2020}]{Byun:2020hun}
Byun J.,  Franco F.~O.,  Howlett C.,  Bonvin C.,   Obreschkow D.,  2020,
  \mn@doi [Mon. Not. Roy. Astron. Soc.] {10.1093/mnras/staa2020}, 497, 1765

\bibitem[\protect\citeauthoryear{Byun, Oddo, Porciani  \& Sefusatti}{Byun
  et~al.}{2021}]{Byun:2020rgl}
Byun J.,  Oddo A.,  Porciani C.,   Sefusatti E.,  2021, \mn@doi [JCAP]
  {10.1088/1475-7516/2021/03/105}, 03, 105

\bibitem[\protect\citeauthoryear{Cabass, Ivanov, Philcox, Simonovi\'c  \&
  Zaldarriaga}{Cabass et~al.}{2022a}]{Cabass:2022ymb}
Cabass G.,  Ivanov M.~M.,  Philcox O. H.~E.,  Simonovi\'c M.,   Zaldarriaga M.,
   2022a

\bibitem[\protect\citeauthoryear{Cabass, Ivanov, Philcox, Simonovi\'c  \&
  Zaldarriaga}{Cabass et~al.}{2022b}]{Cabass:2022wjy}
Cabass G.,  Ivanov M.~M.,  Philcox O. H.~E.,  Simonovi\'c M.,   Zaldarriaga M.,
   2022b

\bibitem[\protect\citeauthoryear{Chan, Scoccimarro  \& Sheth}{Chan
  et~al.}{2012}]{Chan:2012jj}
Chan K.~C.,  Scoccimarro R.,   Sheth R.~K.,  2012, \mn@doi [Phys. Rev. D]
  {10.1103/PhysRevD.85.083509}, 85, 083509

\bibitem[\protect\citeauthoryear{Chartier \& Wandelt}{Chartier \&
  Wandelt}{2021}]{Chartier:2021frd}
Chartier N.,  Wandelt B.~D.,  2021, \mn@doi [Mon. Not. Roy. Astron. Soc.]
  {10.1093/mnras/stab3097}, 509, 2220

\bibitem[\protect\citeauthoryear{Chiang, Wagner, Schmidt  \& Komatsu}{Chiang
  et~al.}{2014}]{Chiang:2014oga}
Chiang C.-T.,  Wagner C.,  Schmidt F.,   Komatsu E.,  2014, \mn@doi [JCAP]
  {10.1088/1475-7516/2014/05/048}, 05, 048

\bibitem[\protect\citeauthoryear{Chiang, Wagner, S\'anchez, Schmidt  \&
  Komatsu}{Chiang et~al.}{2015}]{Chiang:2015eza}
Chiang C.-T.,  Wagner C.,  S\'anchez A.~G.,  Schmidt F.,   Komatsu E.,  2015,
  \mn@doi [JCAP] {10.1088/1475-7516/2015/9/028}, 09, 028

\bibitem[\protect\citeauthoryear{Clarkson, de Weerd, Jolicoeur, Maartens  \&
  Umeh}{Clarkson et~al.}{2019}]{Clarkson:2018dwn}
Clarkson C.,  de Weerd E.~M.,  Jolicoeur S.,  Maartens R.,   Umeh O.,  2019,
  \mn@doi [Mon. Not. Roy. Astron. Soc.] {10.1093/mnrasl/slz066}, 486, L101

\bibitem[\protect\citeauthoryear{Colavincenzo et~al.}{Colavincenzo
  et~al.}{2019}]{Colavincenzo:2018cgf}
Colavincenzo M.,  et~al., 2019, \mn@doi [Mon. Not. Roy. Astron. Soc.]
  {10.1093/mnras/sty2964}, 482, 4883

\bibitem[\protect\citeauthoryear{Dai, Verde  \& Xia}{Dai
  et~al.}{2020}]{Dai:2020adm}
Dai J.-P.,  Verde L.,   Xia J.-Q.,  2020, \mn@doi [JCAP]
  {10.1088/1475-7516/2020/08/007}, 08, 007

\bibitem[\protect\citeauthoryear{Dor\'e et~al.}{Dor\'e
  et~al.}{2014}]{Dore:2014cca}
Dor\'e O.,  et~al., 2014

\bibitem[\protect\citeauthoryear{Eggemeier \& Smith}{Eggemeier \&
  Smith}{2017}]{Eggemeier:2016asq}
Eggemeier A.,  Smith R.~E.,  2017, \mn@doi [Mon. Not. Roy. Astron. Soc.]
  {10.1093/mnras/stw3249}, 466, 2496

\bibitem[\protect\citeauthoryear{Fergusson, Liguori  \& Shellard}{Fergusson
  et~al.}{2010}]{Fergusson:2009nv}
Fergusson J.,  Liguori M.,   Shellard E.,  2010, \mn@doi [Phys. Rev. D]
  {10.1103/PhysRevD.82.023502}, 82, 023502

\bibitem[\protect\citeauthoryear{Fergusson, Liguori  \& Shellard}{Fergusson
  et~al.}{2012a}]{Fergusson:2010dm}
Fergusson J.,  Liguori M.,   Shellard E.,  2012a, \mn@doi [JCAP]
  {10.1088/1475-7516/2012/12/032}, 12, 032

\bibitem[\protect\citeauthoryear{Fergusson, Regan  \& Shellard}{Fergusson
  et~al.}{2012b}]{Fergusson:2010ia}
Fergusson J.~R.,  Regan D.~M.,   Shellard E. P.~S.,  2012b, \mn@doi [Phys.
  Rev.] {10.1103/PhysRevD.86.063511}, D86, 063511

\bibitem[\protect\citeauthoryear{Franco, Bonvin, Obreschkow, Ali  \&
  Byun}{Franco et~al.}{2019}]{Franco:2018yag}
Franco F.~O.,  Bonvin C.,  Obreschkow D.,  Ali K.,   Byun J.,  2019, \mn@doi
  [Phys. Rev.] {10.1103/PhysRevD.99.103530}, D99, 103530

\bibitem[\protect\citeauthoryear{{Friedrich} \& {Eifler}}{{Friedrich} \&
  {Eifler}}{2018}]{FriedrichEifler2018}
{Friedrich} O.,  {Eifler} T.,  2018, \mn@doi [Mon. Not. Roy. Astron. Soc.]
  {10.1093/mnras/stx2566}, \href
  {https://ui.adsabs.harvard.edu/abs/2018MNRAS.473.4150F} {473, 4150}

\bibitem[\protect\citeauthoryear{Gagrani \& Samushia}{Gagrani \&
  Samushia}{2017}]{Gagrani:2016rfy}
Gagrani P.,  Samushia L.,  2017, \mn@doi [Mon. Not. Roy. Astron. Soc.]
  {10.1093/mnras/stx135}, 467, 928

\bibitem[\protect\citeauthoryear{Gil-Marín, Noreña, Verde, Percival, Wagner,
  Manera  \& Schneider}{Gil-Marín et~al.}{2015a}]{Gil-Marin:2014sta}
Gil-Marín H.,  Noreña J.,  Verde L.,  Percival W.~J.,  Wagner C.,  Manera M.,
    Schneider D.~P.,  2015a, \mn@doi [Mon. Not. Roy. Astron. Soc.]
  {10.1093/mnras/stv961}, 451, 539

\bibitem[\protect\citeauthoryear{Gil-Marín et~al.,}{Gil-Marín
  et~al.}{2015b}]{Gil-Marin:2014baa}
Gil-Marín H.,  et~al., 2015b, \mn@doi [Mon. Not. Roy. Astron. Soc.]
  {10.1093/mnras/stv1359}, 452, 1914

\bibitem[\protect\citeauthoryear{Gil-Marín, Percival, Verde, Brownstein,
  Chuang, Kitaura, Rodríguez-Torres  \& Olmstead}{Gil-Marín
  et~al.}{2017}]{Gil-Marin:2016wya}
Gil-Marín H.,  Percival W.~J.,  Verde L.,  Brownstein J.~R.,  Chuang C.-H.,
  Kitaura F.-S.,  Rodríguez-Torres S.~A.,   Olmstead M.~D.,  2017, \mn@doi
  [Mon. Not. Roy. Astron. Soc.] {10.1093/mnras/stw2679}, 465, 1757

\bibitem[\protect\citeauthoryear{Gualdi \& Verde}{Gualdi \&
  Verde}{2020}]{Gualdi:2020ymf}
Gualdi D.,  Verde L.,  2020, \mn@doi [JCAP] {10.1088/1475-7516/2020/06/041},
  06, 041

\bibitem[\protect\citeauthoryear{Gualdi \& Verde}{Gualdi \&
  Verde}{2022}]{Gualdi:2022kwz}
Gualdi D.,  Verde L.,  2022

\bibitem[\protect\citeauthoryear{Gualdi, Manera, Joachimi  \& Lahav}{Gualdi
  et~al.}{2018}]{Gualdi:2017iey}
Gualdi D.,  Manera M.,  Joachimi B.,   Lahav O.,  2018, \mn@doi [Mon. Not. Roy.
  Astron. Soc.] {10.1093/mnras/sty261}, 476, 4045

\bibitem[\protect\citeauthoryear{Gualdi, Gil-Mar\'\i{}n, Manera, Joachimi  \&
  Lahav}{Gualdi et~al.}{2019a}]{Gualdi:2019ybt}
Gualdi D.,  Gil-Mar\'\i{}n H.,  Manera M.,  Joachimi B.,   Lahav O.,  2019a,
  \mn@doi [Mon. Not. Roy. Astron. Soc.] {10.1093/mnrasl/sly242}, 484, L29

\bibitem[\protect\citeauthoryear{Gualdi, Gil-Marín, Schuhmann, Manera,
  Joachimi  \& Lahav}{Gualdi et~al.}{2019b}]{Gualdi:2018pyw}
Gualdi D.,  Gil-Marín H.,  Schuhmann R.~L.,  Manera M.,  Joachimi B.,   Lahav
  O.,  2019b, \mn@doi [Mon. Not. Roy. Astron. Soc.] {10.1093/mnras/stz051},
  484, 3713

\bibitem[\protect\citeauthoryear{Gualdi, Gil-Mar\'\i{}n, Manera, Joachimi  \&
  Lahav}{Gualdi et~al.}{2020}]{Gualdi:2019sfc}
Gualdi D.,  Gil-Mar\'\i{}n H.,  Manera M.,  Joachimi B.,   Lahav O.,  2020,
  \mn@doi [Mon. Not. Roy. Astron. Soc.] {10.1093/mnras/staa1941}, 497, 776

\bibitem[\protect\citeauthoryear{Hahn}{Hahn}{2005}]{Hahn:2004fe}
Hahn T.,  2005, \mn@doi [Comput. Phys. Commun.] {10.1016/j.cpc.2005.01.010},
  168, 78

\bibitem[\protect\citeauthoryear{Hahn}{Hahn}{2015}]{Hahn:2014fua}
Hahn T.,  2015, \mn@doi [J. Phys. Conf. Ser.] {10.1088/1742-6596/608/1/012066},
  608, 012066

\bibitem[\protect\citeauthoryear{Hahn \& Villaescusa-Navarro}{Hahn \&
  Villaescusa-Navarro}{2021}]{Hahn:2020lou}
Hahn C.,  Villaescusa-Navarro F.,  2021, \mn@doi [JCAP]
  {10.1088/1475-7516/2021/04/029}, 04, 029

\bibitem[\protect\citeauthoryear{Hahn, Villaescusa-Navarro, Castorina  \&
  Scoccimarro}{Hahn et~al.}{2020}]{Hahn:2019zob}
Hahn C.,  Villaescusa-Navarro F.,  Castorina E.,   Scoccimarro R.,  2020,
  \mn@doi [JCAP] {10.1088/1475-7516/2020/03/040}, 03, 040

\bibitem[\protect\citeauthoryear{Hajian \& Souradeep}{Hajian \&
  Souradeep}{2003}]{Hajian:2003qq}
Hajian A.,  Souradeep T.,  2003, \mn@doi [Astrophys. J. Lett.]
  {10.1086/379757}, 597, L5

\bibitem[\protect\citeauthoryear{Hajian \& Souradeep}{Hajian \&
  Souradeep}{2005}]{Hajian:2005jh}
Hajian A.,  Souradeep T.,  2005

\bibitem[\protect\citeauthoryear{Hall \& Taylor}{Hall \&
  Taylor}{2019}]{Hall:2018umb}
Hall A.,  Taylor A.,  2019, \mn@doi [Mon. Not. Roy. Astron. Soc.]
  {10.1093/mnras/sty3102}, 483, 189

\bibitem[\protect\citeauthoryear{Heavens, Jimenez  \& Lahav}{Heavens
  et~al.}{2000}]{Heavens:1999am}
Heavens A.,  Jimenez R.,   Lahav O.,  2000, \mn@doi [Mon. Not. Roy. Astron.
  Soc.] {10.1046/j.1365-8711.2000.03692.x}, 317, 965

\bibitem[\protect\citeauthoryear{Hung, Fergusson  \& Shellard}{Hung
  et~al.}{2019a}]{Hung:2019ygc}
Hung J.,  Fergusson J.~R.,   Shellard E. P.~S.,  2019a

\bibitem[\protect\citeauthoryear{Hung, Manera  \& Shellard}{Hung
  et~al.}{2019b}]{Hung:2019nma}
Hung J.,  Manera M.,   Shellard E.,  2019b

\bibitem[\protect\citeauthoryear{Joachimi}{Joachimi}{2017}]{Joachimi:2016xhk}
Joachimi B.,  2017, \mn@doi [Mon. Not. Roy. Astron. Soc.]
  {10.1093/mnrasl/slw240}, 466, L83

\bibitem[\protect\citeauthoryear{Karagiannis, Lazanu, Liguori, Raccanelli,
  Bartolo  \& Verde}{Karagiannis et~al.}{2018}]{Karagiannis:2018jdt}
Karagiannis D.,  Lazanu A.,  Liguori M.,  Raccanelli A.,  Bartolo N.,   Verde
  L.,  2018, \mn@doi [Mon. Not. Roy. Astron. Soc.] {10.1093/mnras/sty1029},
  478, 1341

\bibitem[\protect\citeauthoryear{Lazanu, Giannantonio, Schmittfull  \&
  Shellard}{Lazanu et~al.}{2016}]{Lazanu:2015rta}
Lazanu A.,  Giannantonio T.,  Schmittfull M.,   Shellard E. P.~S.,  2016,
  \mn@doi [Phys. Rev.] {10.1103/PhysRevD.93.083517}, D93, 083517

\bibitem[\protect\citeauthoryear{Lazanu, Giannantonio, Schmittfull  \&
  Shellard}{Lazanu et~al.}{2017}]{Lazanu:2015bqo}
Lazanu A.,  Giannantonio T.,  Schmittfull M.,   Shellard E.,  2017, \mn@doi
  [Phys. Rev. D] {10.1103/PhysRevD.95.083511}, 95, 083511

\bibitem[\protect\citeauthoryear{Lazeyras, Wagner, Baldauf  \&
  Schmidt}{Lazeyras et~al.}{2016}]{Lazeyras:2015lgp}
Lazeyras T.,  Wagner C.,  Baldauf T.,   Schmidt F.,  2016, \mn@doi [JCAP]
  {10.1088/1475-7516/2016/02/018}, 02, 018

\bibitem[\protect\citeauthoryear{Maartens, Jolicoeur, Umeh, De~Weerd, Clarkson
  \& Camera}{Maartens et~al.}{2020}]{Maartens:2019yhx}
Maartens R.,  Jolicoeur S.,  Umeh O.,  De~Weerd E.~M.,  Clarkson C.,   Camera
  S.,  2020, \mn@doi [JCAP] {10.1088/1475-7516/2020/03/065}, 03, 065

\bibitem[\protect\citeauthoryear{Maartens, Jolicoeur, Umeh, De~Weerd  \&
  Clarkson}{Maartens et~al.}{2021}]{Maartens:2020jzf}
Maartens R.,  Jolicoeur S.,  Umeh O.,  De~Weerd E.~M.,   Clarkson C.,  2021,
  \mn@doi [JCAP] {10.1088/1475-7516/2021/04/013}, 04, 013

\bibitem[\protect\citeauthoryear{Moradinezhad~Dizgah, Lee, Schmittfull  \&
  Dvorkin}{Moradinezhad~Dizgah et~al.}{2020}]{MoradinezhadDizgah:2019xun}
Moradinezhad~Dizgah A.,  Lee H.,  Schmittfull M.,   Dvorkin C.,  2020, \mn@doi
  [JCAP] {10.1088/1475-7516/2020/04/011}, 04, 011

\bibitem[\protect\citeauthoryear{Moradinezhad~Dizgah, Biagetti, Sefusatti,
  Desjacques  \& Nore\~na}{Moradinezhad~Dizgah
  et~al.}{2021}]{MoradinezhadDizgah:2020whw}
Moradinezhad~Dizgah A.,  Biagetti M.,  Sefusatti E.,  Desjacques V.,   Nore\~na
  J.,  2021, \mn@doi [JCAP] {10.1088/1475-7516/2021/05/015}, 05, 015

\bibitem[\protect\citeauthoryear{Obreschkow, Power, Bruderer  \&
  Bonvin}{Obreschkow et~al.}{2013}]{Obreschkow:2012yb}
Obreschkow D.,  Power C.,  Bruderer M.,   Bonvin C.,  2013, \mn@doi [Astrophys.
  J.] {10.1088/0004-637X/762/2/115}, 762, 115

\bibitem[\protect\citeauthoryear{Oddo, Rizzo, Sefusatti, Porciani  \&
  Monaco}{Oddo et~al.}{2021}]{Oddo:2021iwq}
Oddo A.,  Rizzo F.,  Sefusatti E.,  Porciani C.,   Monaco P.,  2021, \mn@doi
  [JCAP] {10.1088/1475-7516/2021/11/038}, 11, 038

\bibitem[\protect\citeauthoryear{Pardede, Rizzo, Biagetti, Castorina, Sefusatti
   \& Monaco}{Pardede et~al.}{2022}]{Pardede:2022udo}
Pardede K.,  Rizzo F.,  Biagetti M.,  Castorina E.,  Sefusatti E.,   Monaco P.,
   2022

\bibitem[\protect\citeauthoryear{Pearson \& Samushia}{Pearson \&
  Samushia}{2016}]{Pearson:2015gca}
Pearson D.~W.,  Samushia L.,  2016, \mn@doi [Mon. Not. Roy. Astron. Soc.]
  {10.1093/mnras/stw062}, 457, 993

\bibitem[\protect\citeauthoryear{Philcox}{Philcox}{2021}]{Philcox:2021ukg}
Philcox O. H.~E.,  2021, \mn@doi [Phys. Rev. D] {10.1103/PhysRevD.104.123529},
  104, 123529

\bibitem[\protect\citeauthoryear{Philcox \& Ivanov}{Philcox \&
  Ivanov}{2022}]{Philcox:2021kcw}
Philcox O. H.~E.,  Ivanov M.~M.,  2022, \mn@doi [Phys. Rev. D]
  {10.1103/PhysRevD.105.043517}, 105, 043517

\bibitem[\protect\citeauthoryear{Philcox, Ivanov, Zaldarriaga, Simonovic  \&
  Schmittfull}{Philcox et~al.}{2021}]{Philcox:2020zyp}
Philcox O. H.~E.,  Ivanov M.~M.,  Zaldarriaga M.,  Simonovic M.,   Schmittfull
  M.,  2021, \mn@doi [Phys. Rev. D] {10.1103/PhysRevD.103.043508}, 103, 043508

\bibitem[\protect\citeauthoryear{Pratten \& Munshi}{Pratten \&
  Munshi}{2012}]{Pratten:2011kh}
Pratten G.,  Munshi D.,  2012, \mn@doi [Mon. Not. Roy. Astron. Soc.]
  {10.1111/j.1365-2966.2012.21103.x}, 423, 3209

\bibitem[\protect\citeauthoryear{Regan}{Regan}{2017}]{Regan:2017vgi}
Regan D.,  2017, \mn@doi [JCAP] {10.1088/1475-7516/2017/12/020}, 12, 020

\bibitem[\protect\citeauthoryear{Regan, Schmittfull, Shellard  \&
  Fergusson}{Regan et~al.}{2012}]{Regan:2011zq}
Regan D.~M.,  Schmittfull M.~M.,  Shellard E. P.~S.,   Fergusson J.~R.,  2012,
  \mn@doi [Phys. Rev.] {10.1103/PhysRevD.86.123524}, D86, 123524

\bibitem[\protect\citeauthoryear{Rizzo, Moretti, Pardede, Eggemeier, Oddo,
  Sefusatti, Porciani  \& Monaco}{Rizzo et~al.}{2022}]{Rizzo:2022lmh}
Rizzo F.,  Moretti C.,  Pardede K.,  Eggemeier A.,  Oddo A.,  Sefusatti E.,
  Porciani C.,   Monaco P.,  2022

\bibitem[\protect\citeauthoryear{Ruggeri, Castorina, Carbone  \&
  Sefusatti}{Ruggeri et~al.}{2018}]{Ruggeri:2017dda}
Ruggeri R.,  Castorina E.,  Carbone C.,   Sefusatti E.,  2018, \mn@doi [JCAP]
  {10.1088/1475-7516/2018/03/003}, 03, 003

\bibitem[\protect\citeauthoryear{Saito, Baldauf, Vlah, Seljak, Okumura  \&
  McDonald}{Saito et~al.}{2014}]{Saito:2014qha}
Saito S.,  Baldauf T.,  Vlah Z.,  Seljak U.,  Okumura T.,   McDonald P.,  2014,
  \mn@doi [Phys. Rev. D] {10.1103/PhysRevD.90.123522}, 90, 123522

\bibitem[\protect\citeauthoryear{Samushia, Slepian  \&
  Villaescusa-Navarro}{Samushia et~al.}{2021}]{Samushia:2021ixs}
Samushia L.,  Slepian Z.,   Villaescusa-Navarro F.,  2021, \mn@doi [Mon. Not.
  Roy. Astron. Soc.] {10.1093/mnras/stab1199}, 505, 628

\bibitem[\protect\citeauthoryear{Schmittfull \&
  Moradinezhad~Dizgah}{Schmittfull \&
  Moradinezhad~Dizgah}{2021}]{Schmittfull:2020hoi}
Schmittfull M.,  Moradinezhad~Dizgah A.,  2021, \mn@doi [JCAP]
  {10.1088/1475-7516/2021/03/020}, 03, 020

\bibitem[\protect\citeauthoryear{Schmittfull, Regan  \& Shellard}{Schmittfull
  et~al.}{2013}]{Schmittfull:2012hq}
Schmittfull M.~M.,  Regan D.~M.,   Shellard E. P.~S.,  2013, \mn@doi [Phys.
  Rev.] {10.1103/PhysRevD.88.063512}, D88, 063512

\bibitem[\protect\citeauthoryear{Schmittfull, Baldauf  \& Seljak}{Schmittfull
  et~al.}{2015}]{Schmittfull:2014tca}
Schmittfull M.,  Baldauf T.,   Seljak U.,  2015, \mn@doi [Phys. Rev. D]
  {10.1103/PhysRevD.91.043530}, 91, 043530

\bibitem[\protect\citeauthoryear{Scoccimarro}{Scoccimarro}{2015}]{Scoccimarro:2015bla}
Scoccimarro R.,  2015, \mn@doi [Phys. Rev. D] {10.1103/PhysRevD.92.083532}, 92,
  083532

\bibitem[\protect\citeauthoryear{Scoccimarro, Couchman  \& Frieman}{Scoccimarro
  et~al.}{1999}]{Scoccimarro:1999ed}
Scoccimarro R.,  Couchman H. M.~P.,   Frieman J.~A.,  1999, \mn@doi [Astrophys.
  J.] {10.1086/307220}, 517, 531

\bibitem[\protect\citeauthoryear{Sefusatti, Crocce, Pueblas  \&
  Scoccimarro}{Sefusatti et~al.}{2006}]{Sefusatti:2006pa}
Sefusatti E.,  Crocce M.,  Pueblas S.,   Scoccimarro R.,  2006, \mn@doi [Phys.
  Rev. D] {10.1103/PhysRevD.74.023522}, 74, 023522

\bibitem[\protect\citeauthoryear{Shiraishi, Sugiyama  \& Okumura}{Shiraishi
  et~al.}{2017}]{Shiraishi:2016wec}
Shiraishi M.,  Sugiyama N.~S.,   Okumura T.,  2017, \mn@doi [Phys. Rev. D]
  {10.1103/PhysRevD.95.063508}, 95, 063508

\bibitem[\protect\citeauthoryear{Slepian et~al.}{Slepian
  et~al.}{2017}]{Slepian:2015hca}
Slepian Z.,  et~al., 2017, \mn@doi [Mon. Not. Roy. Astron. Soc.]
  {10.1093/mnras/stw3234}, 468, 1070

\bibitem[\protect\citeauthoryear{Song, Taruya  \& Oka}{Song
  et~al.}{2015}]{Song:2015gca}
Song Y.-S.,  Taruya A.,   Oka A.,  2015, \mn@doi [JCAP]
  {10.1088/1475-7516/2015/08/007}, 08, 007

\bibitem[\protect\citeauthoryear{Sugiyama, Shiraishi  \& Okumura}{Sugiyama
  et~al.}{2018}]{Sugiyama:2017ggb}
Sugiyama N.~S.,  Shiraishi M.,   Okumura T.,  2018, \mn@doi [Mon. Not. Roy.
  Astron. Soc.] {10.1093/mnras/stx2333}, 473, 2737

\bibitem[\protect\citeauthoryear{Sugiyama, Saito, Beutler  \& Seo}{Sugiyama
  et~al.}{2019}]{Sugiyama:2018yzo}
Sugiyama N.~S.,  Saito S.,  Beutler F.,   Seo H.-J.,  2019, \mn@doi [Mon. Not.
  Roy. Astron. Soc.] {10.1093/mnras/sty3249}, 484, 364

\bibitem[\protect\citeauthoryear{Sugiyama, Saito, Beutler  \& Seo}{Sugiyama
  et~al.}{2020}]{Sugiyama:2019ike}
Sugiyama N.~S.,  Saito S.,  Beutler F.,   Seo H.-J.,  2020, \mn@doi [Mon. Not.
  Roy. Astron. Soc.] {10.1093/mnras/staa1940}, 497, 1684

\bibitem[\protect\citeauthoryear{Szapudi}{Szapudi}{2004}]{Szapudi:2004gh}
Szapudi I.,  2004, \mn@doi [Astrophys. J.] {10.1086/423168}, 614, 51

\bibitem[\protect\citeauthoryear{{Tegmark}, {Taylor}  \& {Heavens}}{{Tegmark}
  et~al.}{1997}]{Tegmark1997}
{Tegmark} M.,  {Taylor} A.~N.,   {Heavens} A.~F.,  1997, \mn@doi [Astrophys.
  J.] {10.1086/303939}, \href
  {https://ui.adsabs.harvard.edu/abs/1997ApJ...480...22T} {480, 22}

\bibitem[\protect\citeauthoryear{Tellarini, Ross, Tasinato  \& Wands}{Tellarini
  et~al.}{2016}]{Tellarini:2016sgp}
Tellarini M.,  Ross A.~J.,  Tasinato G.,   Wands D.,  2016, \mn@doi [JCAP]
  {10.1088/1475-7516/2016/06/014}, 06, 014

\bibitem[\protect\citeauthoryear{Varshalovich, Moskalev  \&
  Khersonskii}{Varshalovich et~al.}{1988}]{Varshalovich1988}
Varshalovich D.~A.,  Moskalev A.~N.,   Khersonskii V.~K.,  1988, {Quantum
  Theory of Angular Momentum}.
World Scientific Publishing Co., New Jersey, USA

\bibitem[\protect\citeauthoryear{Wadekar \& Scoccimarro}{Wadekar \&
  Scoccimarro}{2020}]{Wadekar:2019rdu}
Wadekar D.,  Scoccimarro R.,  2020, \mn@doi [Phys. Rev. D]
  {10.1103/PhysRevD.102.123517}, 102, 123517

\bibitem[\protect\citeauthoryear{Wieczorek \& Meschede}{Wieczorek \&
  Meschede}{2018}]{Wieczorek:2018}
Wieczorek M.~A.,  Meschede M.,  2018, \mn@doi [Geochemistry, Geophysics,
  Geosystems] {10.1029/2018GC007529}, 19, 2574

\bibitem[\protect\citeauthoryear{Wolstenhulme, Bonvin  \&
  Obreschkow}{Wolstenhulme et~al.}{2015}]{Wolstenhulme:2014cla}
Wolstenhulme R.,  Bonvin C.,   Obreschkow D.,  2015, \mn@doi [Astrophys. J.]
  {10.1088/0004-637X/804/2/132}, 804, 132

\bibitem[\protect\citeauthoryear{Yamauchi, Yokoyama  \& Tashiro}{Yamauchi
  et~al.}{2017}]{Yamauchi:2017ibz}
Yamauchi D.,  Yokoyama S.,   Tashiro H.,  2017, \mn@doi [Phys. Rev. D]
  {10.1103/PhysRevD.96.123516}, 96, 123516

\bibitem[\protect\citeauthoryear{Yankelevich \& Porciani}{Yankelevich \&
  Porciani}{2019}]{Yankelevich:2018uaz}
Yankelevich V.,  Porciani C.,  2019, \mn@doi [Mon. Not. Roy. Astron. Soc.]
  {10.1093/mnras/sty3143}, 483, 2078

\bibitem[\protect\citeauthoryear{Yankelevich, McCarthy, Kwan, Stafford  \&
  Liu}{Yankelevich et~al.}{2022}]{Yankelevich:2022mus}
Yankelevich V.,  McCarthy I.~G.,  Kwan J.,  Stafford S.~G.,   Liu J.,  2022

\bibitem[\protect\citeauthoryear{de Weerd, Clarkson, Jolicoeur, Maartens  \&
  Umeh}{de~Weerd et~al.}{2020}]{deWeerd:2019cae}
de Weerd E.~M.,  Clarkson C.,  Jolicoeur S.,  Maartens R.,   Umeh O.,  2020,
  \mn@doi [JCAP] {10.1088/1475-7516/2020/05/018}, 05, 018

\makeatother
\end{thebibliography}


\bsp	
\label{lastpage}
\end{document}